\documentclass[journal]{IEEEtranTCOM}
\normalsize
\newif\ifonecolumn
\usepackage{amsmath,amssymb}
\usepackage{float}
\usepackage{placeins}
\usepackage{citesort}
\usepackage{balance}
\usepackage{multirow}
\usepackage{array}
\usepackage{ctable}

\newcommand{\ccolora}{red}
\newcommand{\ccolorb}{blue}
\newcommand{\tcola}[1]{\textcolor{\ccolora}{#1}}
\newcommand{\tcolb}[1]{\textcolor{\ccolorb}{#1}}

\newcommand{\otoprule}{\midrule[\heavyrulewidth]}

\usepackage{algorithm}
\usepackage[noend]{algpseudocode}
\algnewcommand\algorithmicto{\textbf{\ to\ }}

\usepackage{tikz}
\usetikzlibrary{automata,arrows,snakes,shapes}

\renewcommand{\IEEEQED}{\IEEEQEDopen} 

\title{Using Short Synchronous WOM Codes to Make WOM Codes Decodable}

\author{
  Nicolas~Bitouz\'{e},
  Alexandre~Graell~i~Amat,~\IEEEmembership{Senior Member,~IEEE},
  and Eirik~Rosnes,~\IEEEmembership{Senior Member,~IEEE}
  
  \thanks{A.\ Graell i Amat was supported by the Swedish Research Council under Grant \#2011-5961. E.\ Rosnes was supported by Simula@UiB. The material in this paper was presented in part at the 2012
    IEEE International Symposium on Information Theory, Cambridge, MA, July 2012.}
  \thanks{N. Bitouz\'{e} was with the Department
    of Electronics, Institut T\'{e}l\'{e}com-T\'{e}l\'{e}com Bretagne, CS 83818 - 29238
    Brest Cedex 3, France. He is now with the Department of Electrical Engineering,  University of California, Los Angeles (UCLA), 
        Los Angeles, CA 90095-1594. E-mail: bitouze@ucla.edu.}
\thanks{A. Graell i Amat is with the Department of Signals and Systems,
    Chalmers University of Technology, Gothenburg, Sweden. E-mail: alexandre.
    graell@chalmers.se.}
\thanks{E. Rosnes was with Ceragon Networks AS, Kokstadveien 23, N-5257 Kokstad,
    Norway. He is now with the Selmer Center, Department of Informatics, University of Bergen, N-5020 Bergen, Norway, and the Simula Research Lab. E-mail: eirik@ii.uib.no.}
}

\DeclareMathOperator{\Image}{Im}

\newtheorem{theorem}{Theorem}

\newtheorem{definition}{Definition}
\newtheorem{proposition}{Proposition}
\newtheorem{example}{Example}

\newcommand{\deltaequals}{\stackrel{\Delta}{=}}
\newcommand{\calD}{\mathcal{D}}
\newcommand{\calE}{\mathcal{E}}

\newcommand{\boldb}{\mathbf{b}}

\newcommand{\bolde}{\mathbf{e}}
\newcommand{\boldx}{\mathbf{x}}
\newcommand{\boldy}{\mathbf{y}}
\newcommand{\sync}{^{\mathrm{sync}}}

\newcommand{\Rnondec}{R_{\mathrm{nd}}}
\newcommand{\nsync}{n_{\rm sync}}
\newcommand{\nnd}{n_{\rm nd}}

\newcommand{\tnd}{t_{\rm nd}}

\begin{document}
\maketitle


\begin{abstract}
In the framework of write-once memory (WOM) codes, it is important to distinguish between codes that can be decoded directly and those that require that the decoder knows the current generation to successfully decode the state of the memory. A widely used approach to construct WOM codes is to design first nondecodable codes that approach the boundaries of the capacity region, and then make them decodable by appending additional cells that store the current generation, at an expense of a rate loss. In this paper, we propose an alternative method to make nondecodable WOM codes decodable by appending cells that also store some additional data. The key idea is to append to the original (nondecodable) code a short \emph{synchronous} WOM code and write generations of the original code and of the synchronous code simultaneously. We consider both the binary and the nonbinary case. Furthermore, we propose a construction of synchronous WOM codes, which are then used to make nondecodable codes decodable. For short-to-moderate block lengths, the proposed method significantly reduces the rate loss as compared to the standard method.
\end{abstract}



\section{Introduction and Definitions}

The write-once memory (WOM) model was introduced in \cite{RivSha82} to study storage devices consisting of $q$-ary ($q\geq 2$) memory cells whose values cannot be decreased. It was originally introduced to model the behavior of optical disks and study coding schemes that would allow one to write data several times on a disk even though each bit can only be written once. By allowing data from a previous write to be ``forgotten'' when a new write occurs, one can show that the total amount of information that can be stored on such a disk is greater if several small pieces of information are stored and forgotten one after the other than if the whole disk is written at once. The model is now mainly studied because of its similarity with flash memories, on which the value of a cell can be decreased, but at an extremely high cost. Since the original paper by Rivest and Shamir \cite{RivSha82}, several other works on this topic have appeared, both in terms of code constructions, capacity, and error-correction. See, for instance, \cite{mer84,fia84,capacitypermanentmemory,cho86,zem91,capacitygeneralizedWOM,yaa10,twowritewomcodes,multiplewritewomcodes,nonbinarywomcodes,yaa12_1,shp13} and references therein. Recently, lattice-based constructions have been proposed. For instance, in \cite{bha14,bha12} lattice-based $t$-write codes for multilevel cells were presented. For applications to flash memories, see \cite{jia07,jia08,mah09}.

The fundamental problem in the WOM model is, considering an array of $n$ empty $q$-ary cells, to know how much information one can store using exactly $t$ writes (also called \emph{generations}). The coding schemes that are used to fulfill this goal are called \emph{$t$-write WOM codes}. The following definition is taken from \cite{nonbinarywomcodes}.
\begin{definition}
 An $[n,t:M_1,\dots,M_t]_q$ $t$-write $q$-ary WOM code $C$ is a coding scheme for $n$ $q$-ary WOM cells, which consists of $t$ pairs of encoding and decoding mappings $\calE_i$ and $\calD_i$ ($1\leq i\leq t$) such that
\ifonecolumn
\else
    \begin{enumerate}
    \item $\calE_1:\{1,\dots,M_1\}\to \{0,\dots,q-1\}^n$.
    \item For $2\leq i \leq t$:
      \begin{itemize}
      \item $\calE_i:\{1,\dots,M_i\}\times \Image(\calE_{i-1})\to \{0,\dots,q-1\}^n$,
      \item $\forall(m,\boldb)\in\{1,\dots,M_i\}\times \Image(\calE_{i-1})$,\\
        $\forall j\in\{1,\dots,n\},\,(\calE_i(m,\boldb))_j\geq (\boldb)_j$.
      \end{itemize}
    \item For $1\leq i\leq t$, $\calD_i:\{0,\dots,q-1\}^n\to\{1,\dots,M_i\}$, and
      \begin{itemize}
      \item $\forall m\in\{1,\dots,M_1\}$, $\calD_1(\calE_1(m))=m$,
      \item for $2\leq i \leq t$, $\forall (m,\boldb)\in\{1,\dots,M_i\}\times \Image(\calE_{i-1})$, $\calD_i(\calE_i(m,\boldb))=m$.
      \end{itemize}
    \end{enumerate}
\fi
\end{definition}

For simplicity, in the remainder of the paper, we will refer to WOM codes simply as codes. The rate of the above code, referred to as the WOM-rate, or sometimes just as the rate of the code, is defined as follows \cite{nonbinarywomcodes}.
\begin{definition}
The rate of generation $i\in\{1,\dots,t\}$ of an $[n,t:M_1,\dots,M_t]_q$ $q$-ary code $C$ is
\begin{equation} \notag
  R_i(C)\deltaequals \frac{\log_2 M_i}{n}
\end{equation}
and the WOM-rate of $C$ is defined as
\begin{equation} \notag
  R(C)\deltaequals \sum_{i=1}^tR_i(C)=\frac{\sum_{i=1}^t\log_2 M_i}{n}.
\end{equation}
\end{definition}
The fundamental problem of the WOM model is therefore to find a code of maximum WOM-rate given $t$ and $q$, and sometimes $n$. 

For some codes, the state of the cells is enough to determine the current generation (i.e., how many times the memory has been written). However, some codes have a structure such that the same state of the memory can appear at different generations. This is not a problem if the same state of the memory at different generations corresponds to the same message, but when it is not the case, the decoder has to be given the knowledge of the current generation in order to successfully decode the memory. We say that a code is \emph{decodable} if for any state of the cells $\boldb$ and any $i_1$ and $i_2$ with $\boldb\in \Image(\calE_{i_1})\cap \Image(\calE_{i_2})$, $\calD_{i_1}(\boldb)=\calD_{i_2}(\boldb)$. A code that does not satisfy this property is called \textit{nondecodable}. A stronger property is given in \cite{RivSha82}: a code is called \emph{synchronous}\footnote{Our concept of a synchronous code is equivalent to the concept of an \emph{almost-synchronous} code from \cite{RivSha82}.} if the current state of the memory provides enough information to know the current generation, i.e., the sets $\Image(\calE_i)$ are disjoint for $1\leq i\leq t$. Synchronous codes are decodable. However, the reverse does not always hold. The work in \cite{RivSha82} also considers a way to guarantee synchronousness: \emph{laminar} codes are codes such that the \emph{weight} of the cells, defined as the $\ell_1$-norm of the $q$-ary cell vector, is an injective function of the generation, i.e., for $\boldb_1\in \Image(\calE_{i_1})$ and $\boldb_2\in \Image(\calE_{i_2})$, $w(\boldb_1)=w(\boldb_2)\Rightarrow i_1=i_2$. In the binary case, the weight reduces to the standard Hamming weight. The authors of \cite{RivSha82} give a construction of laminar codes for $n=t$ being a power of two, with WOM-rate $\log_2(t)/2$. However, synchronous codes have not been extensively studied in the literature. Note that nonsynchronous codes can still be directly decoded if, when the decoder cannot determine the current generation, the choice of $\calD_i$ has no impact on the decoded symbol. In Section~\ref{sec:MainIdea}, we give examples of laminar, synchronous (but nonlaminar), and decodable (but nonsynchronous) codes.

A nondecodable $[\nnd,\tnd:M_1,\dots,M_{\tnd}]_2$ binary code $C$ can be made decodable (and even synchronous) by simply concatenating $k$ instances of $C$ with a block of $\tnd-1$ cells that store the current generation (by being filled one by one at each write, starting at the second generation). The resulting code is a synchronous code with parameters $[k\nnd+\tnd-1,\tnd:M_1^k,\dots,M_{\tnd}^k]_2$. As $k$ goes to infinity, the WOM-rate of this code approaches the WOM-rate of the original code, $R(C)$. 

Most of the state-of-the-art high-rate codes are not \emph{directly} decodable. Indeed, a common approach in the literature is to design (nondecodable) codes that approach the boundaries of the capacity region (see, e.g., \cite{yaa12_1,nonbinarywomcodes}), and then make them decodable using the method above. However, for short-to-moderate block lengths, making a nondecodable code decodable by appending $\tnd-1$ cells containing no data can significantly degrade its WOM-rate. For instance, consider $n=6$ and $t=4$, and assume that we do not know a decodable code of length $6$. In this case, we could select a nondecodable $4$-write code of length $3$, and append $3$ cells to store the current generation. The resulting WOM-rate is half the original one, as the additional cells only carry information about the current generation. 

In this paper, we propose a different approach to make a nondecodable $\tnd$-write code $C$ decodable. Our main focus is on binary codes, but we also extend our results to $q>2$. The key idea is to append (for a $\tnd$-write nondecodable binary code of length $\nnd$) $\tnd-1$ additional cells which store not only the current generation but also new data, by using a $\tnd$-write synchronous code with length $\tnd-1$, and writing generations of $C$ and of the synchronous code simultaneously. Since synchronous codes are at the basis of the proposed method, we consider first the construction of synchronous codes. Our main focus is on laminar codes. The construction of synchronous (laminar) codes was already addressed in \cite{RivSha82}. However, \cite{RivSha82} only considered the case where $n=t$ and $t$ is a power of $2$. Here, we construct small laminar codes for both $n=t$ and $n>t$, and propose a construction for synchronous codes of higher values of $t$. Lifting the constraint $n=t$ allows to achieve higher WOM-rates. The obtained codes are then used to make nondecodable codes decodable. Whereas the main focus of this paper is on \textit{unrestricted-rate} codes \cite{yaa12_1}, i.e., we allow the individual writes to use a different number of inputs, we also extend our construction to \textit{fixed-rate} codes, i.e., codes for which all writes store the same number of messages.

The remainder of this paper is organized as follows. In Section~\ref{sec:MainIdea}, we introduce the main idea to turn nondecodable codes into decodable ones, and provide some examples. In Section~\ref{sec:w(imi)=i}, we consider a simple family of laminar codes with $n=t$, as well as very short codes from this family. We also give bounds on the sizes of their generations, and construct better laminar codes with $n>t$ by local manipulations of the codes with $n=t$. In Section~\ref{sec:construction}, we propose a construction of synchronous codes with good properties to reach higher values of $t$ by concatenating instances of a synchronous code using a second synchronous code to decide, at each generation, which of the instances of the first code are going to be modified. In Section~\ref{sec:fixed_rate}, we study the case of fixed-rate codes, and we extend our results on the binary case to nonbinary scenarios in Section~\ref{sec:qary}. 
Finally, in Section~\ref{sec:results}, we compare our method of making nondecodable codes decodable with the method that only adds 
cells containing no data. Some conclusions are drawn in Section~\ref{sec:conclu}.

\section{Main Idea and Examples}
\label{sec:MainIdea}

Let $C$  be a nondecodable code with parameters $[\nnd,\tnd:M_1,\dots,M_{\tnd}]_2$, and WOM-rate $\Rnondec$. The standard approach to turn $C$ into a decodable code is to append $\tnd-1$ cells that store the current generation, thus obtaining a code of length $n=\nnd+\tnd-1$. This incurs a rate loss 
\begin{align}
\label{eq:Gbasic}
\gamma_{\rm basic} = \frac{\Rnondec - \Rnondec  \frac{\nnd}{n} }{\Rnondec} = \frac{\tnd-1}{n}.
\end{align}%

The main idea in this paper is very simple: instead of adding cells that do not contain information, we append to the original code cells that also store actual data. This is achieved by appending to $C$ a $\tnd$-write synchronous code of length $\nsync=\tnd-1$, and writing generations of $C$ and of the synchronous code simultaneously. Appending a synchronous code to $C$ results in an overall decodable (and also synchronous) code (the synchronousness of the appended code guarantees that by observing the $\tnd-1$ new cells, the decoder can always determine the current generation, and use this knowledge to decode the overall code), while allowing to store extra data.

Let $R_{\rm sync}>0$ be the WOM-rate of the synchronous code that we append to the nondecodable code. The rate loss introduced by this method, denoted by $\gamma_{\rm sync}$, is
\ifonecolumn
\else
\begin{align}
\label{eq:Gsync}
\gamma_{\rm sync} &= \frac{\Rnondec - \left(\Rnondec  (n-\nsync) + R_{\rm sync}  \nsync \right)/n }{\Rnondec}\nonumber\\
& = \frac{\nsync}{n} \left(1-\frac{ R_{\rm sync} }{\Rnondec}\right)
\end{align}
\fi
which is smaller than $\gamma_{\rm basic}$, since we can choose $\nsync = \tnd-1$ (or slightly above). Note that $\gamma_{\rm sync}$ is decreasing with $R_{\rm sync}$ when $n$, $\nsync$, and $\Rnondec>0$ are fixed. The main ingredient of the proposed technique is therefore a $\tnd$-write synchronous code of length $\tnd-1$. To increase $R_{\rm sync}$ one may also consider synchronous codes with $\nsync$ slightly larger than $\tnd-1$ (the length of the resulting overall code would be slightly larger than that of the code obtained applying the standard method. However, the increase in length is compensated by a larger WOM-rate  $R_{\rm sync}$).

The following sections are devoted to the construction of $t$-write synchronous codes of length $n=t-1$ (or slightly larger) to be used to make a nondecodable code decodable as explained above. Ideally, we would like to design synchronous codes that maximize the WOM-rate. However, this is overly complex. Instead we first construct small laminar codes, and then propose a construction method to construct synchronous codes for larger values of $t$ by concatenating smaller codes. The use of laminar codes makes the computer search more tractable.

The construction method in Section~\ref{sec:construction} requires component codes which do not contain the all-zero codeword. Therefore, in Section~\ref{sec:w(imi)=i} we construct small laminar codes which do not contain the all-zero codeword. Note that for codes that do not contain the all-zero codeword, the number of writes is limited by the code length, $t\le n$. Thus, our approach is to construct $(t-1)$-write synchronous codes with length $n=t-1$ from component codes which do not contain the all-zero codeword, and then obtain a $t$-write synchronous code with length $n=t-1$ by simply adding a generation that only contains the all-zero codeword.

To ease the understanding of the paper, in the following we clarify this and the concepts of synchronous, laminar, and decodable (but not synchronous) codes with some examples. For later use, if an $[n,t:M_1,\dots,M_t]_q$ code is synchronous, 
we will frequently use the superscript ``$\mathrm{sync}$'',~$[n,t:M_1,\dots,M_t]_q\sync$. 
Also, in the binary case, the cells that can be written from $0$ to $1$ but not from $1$ to $0$ are called \emph{wits} \cite{RivSha82}.
\begin{example} \label{ex:laminar}
An example of a binary $[4,4:4,2,2,1]_2$ laminar code is depicted in Fig.~\ref{fig:example_laminar} by a state diagram describing all four writes. The four-bit vector in each state is the memory-state. The different types of edges (solid, dashed, dotted, and dash-dotted) correspond to different input data bits. As can be seen from the figure, the weight of the cells uniquely identifies the generation. 
\end{example}

\ifonecolumn
\else
\begin{figure}[!ht]
\par
\begin{center}
\includegraphics[width=\columnwidth]{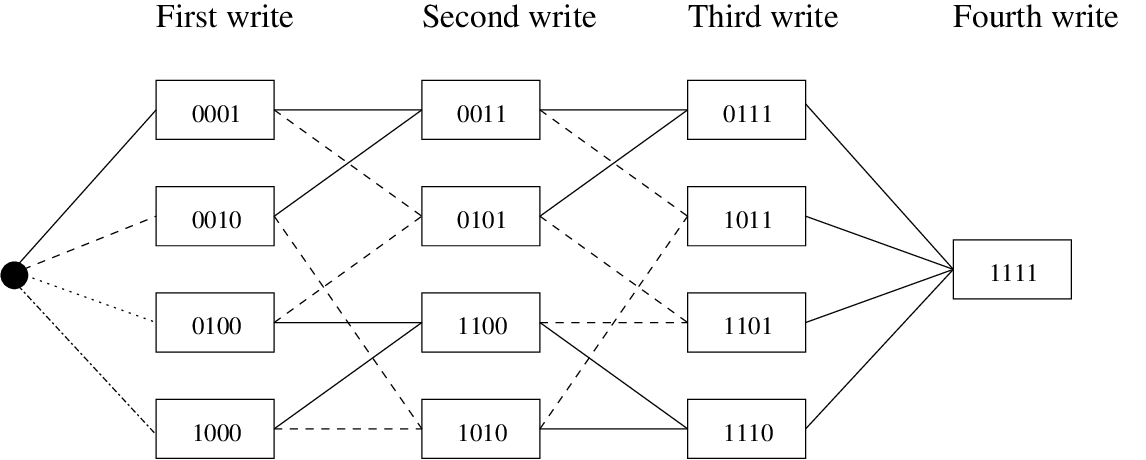}
\end{center}
\caption{\label{fig:example_laminar} {A binary $[4,4:4,2,2,1]_2$ laminar four-write code. The different types of edges (solid, dashed, dotted, and dash-dotted) correspond to different input data bits.}}
 \end{figure}
\fi
\begin{example} \label{ex:sync_not_laminar}
An example of a quaternary $[2,4:2,2,3,3]_4$ synchronous (but nonlaminar)  code is depicted in Fig.~\ref{fig:example_sync_nonlaminar} by a state diagram describing all four writes. The two-symbol vector in each state is the memory-state. The different types of edges (solid, dashed, and dotted) correspond to different input data symbols. As can be seen from the figure, the cells of the memory cannot be in the same state at different generations, which implies that the code is synchronous, but the weight (or $\ell_1$-norm) of the cell state $(22)$ of the third generation and the weight of the cell state $(31)$ (or $(13)$) of the fourth generation are the same. Thus, the weight is \emph{not} an injective function of the generation, and the code is not laminar.
\end{example}
\ifonecolumn
\else
\begin{figure}[!ht]
\par
\begin{center}
\includegraphics[width=\columnwidth]{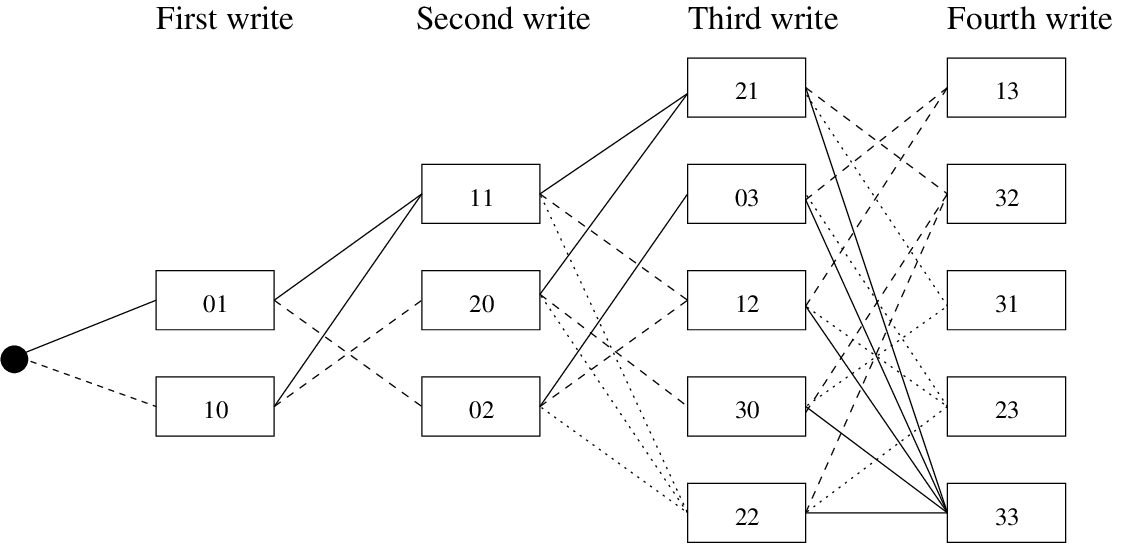}
\end{center}
\caption{\label{fig:example_sync_nonlaminar} {A quaternary $[2,4:2,2,3,3]_4$ synchronous (but nonlaminar) four-write code. The different types of edges (solid, dashed, and dotted) correspond to different input data symbols.}}
 \end{figure}
\fi
\begin{example}
A simple example of a decodable (but nonsynchronous) binary code,  taken from \cite{RivSha82},  that enables two bits to be written into three memory cells twice, is given in Table~\ref{table:nonsync}, which describes the encoding and decoding rules for the code. The code is nonsynchronous, since for the second write, if the information to be encoded does not change, then the state of the memory does not change either. Thus, the current state of the memory does not provide enough information to tell the current generation.
\end{example}
\begin{table}[!ht]
\footnotesize
  \centering
  \caption{A binary $[3,2:4,4]_2$  decodable (but nonsynchronous) code.}
\begin{tabular}{ccc}
\toprule 
Data bits & First write & Second write (if data changes) \\ 
\otoprule
00 & 000 &  111 \\ [0.5mm]
10 & 100 &  011\\ [0.5mm]
01 & 010 &  101\\ [0.5mm]
11 & 001 & 110 \\[0.5mm]
\midrule
\bottomrule
\end{tabular}
  \label{table:nonsync}
\end{table}
\begin{example}
By adding a generation containing the all-zero codeword  prior to all other generations of the  $[4,4:4,2,2,1]_2$  code from Example~\ref{ex:laminar} (and depicted in Fig.~\ref{fig:example_laminar}), the code is turned into a $[4,5:1,4,2,2,1]_2$ code. The WOM-rate is the same, but the number of writes is now the length plus one. The code is depicted in Fig.~\ref{fig:example_laminar_with}.
\end{example}
\ifonecolumn
\else
\begin{figure}[!ht]
\par
\begin{center}
\includegraphics[width=\columnwidth]{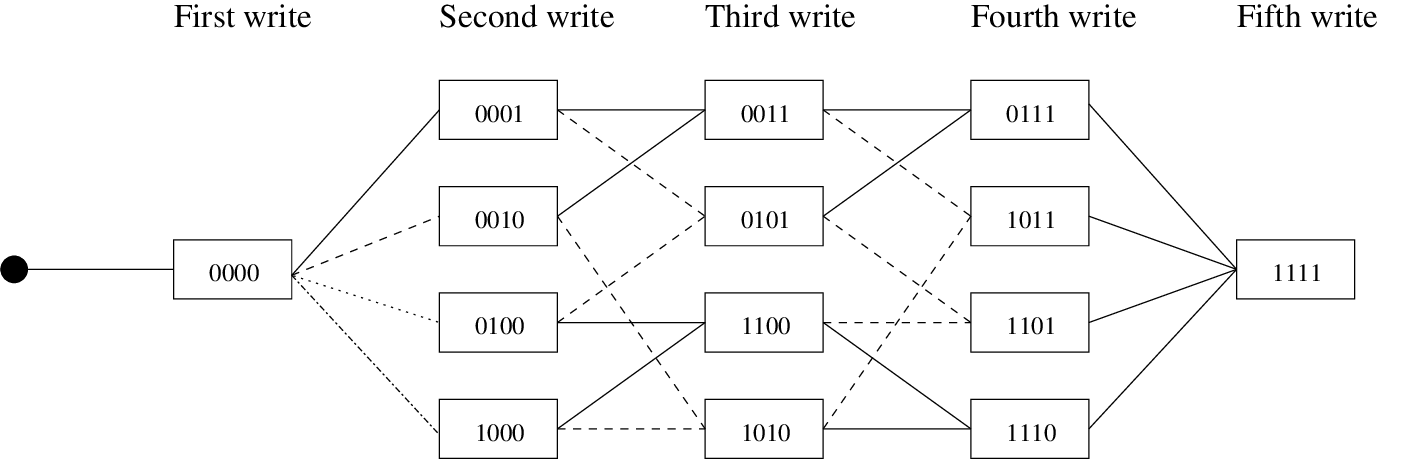}
\end{center}
\caption{\label{fig:example_laminar_with} {A binary $[4,5:1,4,2,2,1]_2$ code obtained from the code of Example~\ref{ex:laminar} by adding a generation prior to all other generations containing the all-zero codeword only. The different types of edges (solid, dashed, dotted, and dash-dotted) correspond to different input data bits.}}
 \end{figure}
\fi

\section{Small Laminar WOM Codes}
\label{sec:w(imi)=i}

In this section, we construct small laminar codes. We first consider codes with $n=t$ that write exactly $1$ wit at each generation, and then construct codes with $n>t$.

An exhaustive search for laminar codes that maximize the WOM-rate is unfeasible even for very short codes. Thus, to simplify the search, we use a greedy algorithm that maximizes the values of $M_i$ generation by generation. Consider a code $C$ with $n=t$ that writes exactly $1$ wit per generation, and a generation $i>1$. Assuming that the previous generations are already fixed, the condition we have on $M_i$ is that for every $\boldx\in \Image(\calE_{i-1})$, and for every $m\in\{1,\dots,M_i\}$, there exists $\boldy\in \Image(\calE_i)$ such that $\boldx\leq\boldy$ and $\calD_i(\boldy)=m$ (where $\boldx\leq\boldy$ if $x_k\leq y_k$ for all $k$, $1\leq k\leq n$). Denote by $E(n,i)$ the set of binary vectors of length $n$ and Hamming weight $i$. It follows that at each generation $i$, $\Image(\calE_i)\subseteq E(n,i)$. We use this set inclusion to make our maximization at each generation completely independent from the other generations, at the cost of optimality.

Let us define the equivalence relation $\equiv_i^n$ on $\Image(\calE_i)$ by $\boldy\equiv_i^n \boldy'$ if and only if $\calD_i(\boldy)=\calD_i(\boldy')$. Let us refer to the equivalence classes of this relation as the \emph{codeword classes} of $C$ at generation $i$. Codeword classes are subsets $Y\subseteq E(n,i)$ for which, if we do not take the previous generations into account, the following must hold
\begin{equation}
  \label{eqn:Y}
  \forall \boldx\in E(n,i-1),\,\exists \boldy\in Y:\,\boldx\leq \boldy.
\end{equation}

We are also interested in the partitions of $E(n,i)$ as a set of valid codeword classes. If $\mathcal{Y}$ denotes such a partition, we want that
\begin{equation}
  \label{eqn:partition}
  \forall Y\in\mathcal{Y},\,\forall \boldx\in E(n,i-1),\,\exists \boldy\in Y:\,\boldx\leq \boldy.
\end{equation}
Each valid partition $\mathcal{Y}$ corresponds to a valid decoding mapping (modulo reordering), and thus each cardinality $|\mathcal{Y}|$ to a valid $M_i$. We are therefore interested in finding the maximum cardinality of such a partition. We make the following important definition.
\begin{definition}
Let $A(n,i)$ be the maximum cardinality of a partition $\mathcal{Y}$ of $E(n,i)$ satisfying (\ref{eqn:partition}).
\end{definition}

We now give an upper bound on $A(n,i)$.

\begin{proposition}
Let $B(n,i)$ be defined by
\begin{equation} \notag
  B(n,i) \deltaequals  \left \lfloor  \frac{{n\choose i}}{\displaystyle\min_{Y\text{ s.t. (\ref{eqn:Y}) holds}}|Y|} \right\rfloor.
\end{equation}
Then, the maximum cardinality $A(n,i)$ of a partition $\mathcal{Y}$ that satisfies (\ref{eqn:partition}) is upper-bounded by $A(n,i)\leq B(n,i)$.
\end{proposition}

\begin{IEEEproof}
  Let $\mathcal{Y}$ be any partition of $E(n,i)$. Then,
  \begin{equation} \notag
    |\mathcal{Y}|\cdot\left(\displaystyle\min_{Y\text{ s.t. (\ref{eqn:Y}) holds}}|Y| \right)\leq   \sum_{Y\in\mathcal{Y}}|Y|=|E(n,i)|={n\choose i}.
  \end{equation}
  This holds in particular when $\mathcal{Y}$ is of maximum cardinality.
\end{IEEEproof}

\begin{table*}[!ht]
\footnotesize
  \centering
  \caption{Upper bound $B(n,i)$ on $A(n,i)$. Values in bold are exact values for $A(n,i)$ found by computer search ($A(n,i)=B(n,i)$ in all cases). The values for $B(n,i)$ in italics match the exact values of $A(n,i)$ by Propositions~\ref{prop:aqn1}, \ref{prop:a22^n2}, and \ref{prop:2n+1}, or by the lower bounds of Propositions~\ref{prop:aqni_vs_aqn-1} and \ref{prop:a22n2}.}
  \begin{tabular}{r|rrrrrrrrrrrrrrrrr}
~~~$i$&     1   &       2  &       3  &       4  &      5  &      6  &      7  &      8  &      9  &     10  &     11  &     12  &     13  &     14  &     15  &   16  \\
$n$~~~&         &          &          &          &         &         &         &         &         &         &         &         &         &         &         &       \\ \hline
1   &{\bf \em 1} &          &          &          &         &         &         &         &         &         &         &         &         &         &         &       \\
2   &{\bf \em 2} & {\bf  \em 1} &          &          &         &         &         &         &         &         &         &         &         &         &         &       \\
3   &{\bf \em 3} & {\bf  \em 1} &  {\bf  1 }&          &         &         &         &         &         &         &         &         &         &         &         &       \\
4   &{\bf \em 4} & {\bf  \em 3} &  {\bf \em 1 }&  {\bf  1 }&         &         &         &         &         &         &         &         &         &         &         &       \\
5   &{\bf \em 5} & {\bf  \em 3} &  {\bf 2 }&  {\bf \em 1 }& {\bf  1 }&         &         &         &         &         &         &         &         &         &         &       \\
6   &{\bf \em 6} & {\bf  5} &       3  &       2  & {\bf \em 1 }& {\bf  1 }&         &         &         &         &         &         &         &         &         &       \\
7   &{\bf \em 7} & {\bf \em 5} &       5  &       2  &      2  & {\bf \em 1 }& {\bf  1 }&         &         &         &         &         &         &         &         &       \\
8   &{\bf \em 8} & {\em  7}  &       5  &       5  &      2  &      2  & {\bf \em 1 }& {\bf 1 }&         &         &         &         &         &         &         &       \\
9   &{\bf \em 9} & {\em  7}  &       6  &       5  &      3  &      2  &      2  & {\bf \em 1 }& {\bf  1 }&         &         &         &         &         &         &       \\
10  &{\bf \em 10} &       9  &       6  &       5  &      4  &      3  &      2  &      2  & {\bf \em 1 }& {\bf  1 }&         &         &         &         &         &       \\
11  &{\bf \em 11} &       9  &       7  &       6  &      5  &      4  &      3  &      2  &      2  & {\bf \em 1 }& {\bf  1 }&         &         &         &         &       \\
12  &{\bf \em 12} & {\em 11}  &       8  &       6  &      6  &      5  &      3  &      3  &      2  &   {\em   1}  & {\bf \em 1 }& {\bf  1 }&         &         &         &       \\
13  &{\bf \em 13} &  {\em 11}  &      10  &       7  &      6  &      5  &      4  &      3  &      3  &      2  &      2  & {\bf \em 1 }& {\bf 1 }&         &         &       \\
14  &{\bf \em 14} &      13  &      10  &       9  &      7  &      6  &      5  &      5  &      3  &      3  &      2  &      2  & {\bf \em 1 }& {\bf  1 }&         &       \\
15  &{\bf \em 15} &      13  &      13  &       9  &      9  &      6  &      5  &      5  &      4  &      3  &      3  &      2  &      2  & {\bf \em 1 }&{\bf   1} &       \\
16  &{\bf \em 16} & {\em 15}  &      13  &      13  &      9  &      9  &      7  &      6  &      5  &      4  &      3  &      3  &      2  &      2  & {\bf \em 1 }&{\bf  1}
  \end{tabular}
  \label{table:bound_2}
\end{table*}

This bound can be computed using a computer search for the smallest $Y$ that satisfies (\ref{eqn:Y}). The search is relatively slow, but notice that by lower-bounding $|Y|$ by $\left\lceil \frac{|E(n,i-1)|}{i} \right\rceil$ (each element $\boldy\in E(n,i)$ covers exactly $i$ elements $\boldx\in E(n,i-1)$), we obtain a closed-form bound,
\begin{equation} \notag
  A(n,i)\leq B(n,i) \leq \left \lfloor  \frac{{n\choose i}}{\left\lceil \frac{|E(n,i-1)|}{i}\right\rceil} \right\rfloor = \left \lfloor  \frac{{n\choose i}}{\left\lceil \frac{{n \choose i-1}}{i}\right\rceil} \right\rfloor.
\end{equation}

While the closed-form bound can be computed efficiently and is reached for some values of $(n,i)$ (for instance, for $n\leq 3$, or for $i\leq 2$, or $i=n$), even for relatively low values of $n$ and $i$, it can be strictly higher than $A(n,i)$. For instance, $A(4,3)=1$, while the closed-form bound is $2$. Indeed, $E(4,3)=\{1110,1101,1011,0111\}$ and $E(4,2)=\{1100,1010,1001,0110,0101,0011\}$, and while each element of $E(4,3)$ covers $3$ elements of $E(4,2)$, it is not possible to pick two elements of $E(4,3)$ such that the subsets of $E(4,2)$ that they cover are disjoint. Therefore, the codeword classes in $E(4,3)$ have cardinality at least $3$, and not $\frac{|E(n,i-1)|}{i}=2$.

For very small values of $n$, the exact value of $A(n,i)$ can be computed by conducting a simple exhaustive search on the set of codeword classes. Values of $B(n,i)$ are also obtained with an exhaustive search, but on the minimum size of codeword classes, which is significantly faster. The results of the two searches are reported for $n\leq 16$ in Table~\ref{table:bound_2}. The values in bold font are $A(n,i)$, the others are $B(n,i)$. The few values of $A(n,i)$ that were computed exactly match $B(n,i)$, so it is unknown whether there are pairs $(n,i)$ such that $A(n,i)<B(n,i)$. Note that these values are constructive. For instance,  a $[4,4:4,3,1,1]_2\sync$ and a $[5,5:5,3,2,1,1]_2\sync$ code can be obtained from the search. The upper bounds from Table~\ref{table:bound_2} in italics match the exact values of $A(n,i)$ by Propositions~\ref{prop:aqn1}, \ref{prop:a22^n2}, and \ref{prop:2n+1}, or by the lower bounds of Propositions~\ref{prop:aqni_vs_aqn-1} and \ref{prop:a22n2}, and are also constructive (see Section~\ref{sec:proofs} below).

\subsection{Bounds on the Sizes of Generations}
\label{sec:proofs}

We give bounds on the sizes of the generations of the codes defined above. In particular, we give lower bounds that are constructive and allow us to effectively build codeword classes for the corresponding generations.

For $\boldx=(x_1,\dots,x_n)\in \{0,1\}^n$ and $\boldx'=(x'_1,\dots,x'_n)\in\{0,1\}^{n'}$, we denote by $\boldx\cdot \boldx'$ the vector of $\{0,1\}^{n+n'}$ that is the concatenation of $\boldx$ and $\boldx'$:
\begin{equation} \notag
  \boldx\cdot \boldx'=(x_1,\dots,x_n,x'_1,\dots,x'_n).
\end{equation}

We also call $\mathcal{Y}$ a \emph{suitable partition} of $E(n,i)$ if (\ref{eqn:partition}) holds, and we do not mind if the union of the elements of $\mathcal{Y}$ is only a strict subset of $E(n,i)$.

\begin{proposition}
  \label{prop:aqni_vs_aqn-1}
  For any $n\geq 2$ and $2\leq i\leq n$, $A(n,i)\geq \min(A(n-1,i-1),A(n-1,i))$.
\end{proposition}

\begin{IEEEproof}
  Let $\mathcal{Y}$ be a suitable partition of $E(n-1,i)$ and $\mathcal{Z}$ a suitable partition of $E(n-1,i-1)$ such that $|\mathcal{Y}|=A(n-1,i)$ and $|\mathcal{Z}|=A(n-1,i-1)$. Consider two bijections $f_{\mathcal{Y}}:\{1,\dots,A(n-1,i)\}\to \mathcal{Y}$ and $f_{\mathcal{Z}}:\{1,\dots,A(n-1,i-1)\}\to \mathcal{Z}$. Now, define a suitable partition $\mathcal{Y}'$ of $E(n,i)$ as the union for all 
$1\leq k\leq \min(A(n-1,i-1),A(n-1,i))$ 
of the codeword classes
\begin{equation} \notag
  \left(f_{\mathcal{Y}}(k).0\right)\,\cup\,\left(f_{\mathcal{Z}}(k).1\right).
\end{equation}
There is no collision between these codeword classes, since we can sort their elements according to their last symbol, and for a given last symbol, the first $n-1$ symbols of the codewords in a codeword class match a (suitable) partition of $E(n-1,i)$ or one of $E(n-1,i-1)$. The cardinality of $\mathcal{Y}'$ is $\min(A(n-1,i-1),A(n-1,i))$.
\end{IEEEproof}

\begin{proposition}
  \label{prop:a22n2}
  For $n\geq 1$, $A(2n,2)\geq 2A(n,2)+1$.
\end{proposition}

\begin{IEEEproof}
  Let $\mathcal{Y}$ be a partition of $E(n,2)$ with cardinality $A(n,2)$ such that for all $Y\in\mathcal{Y}$ and for all $\boldx\in E(n,1)$, there is $\boldy\in Y$ such that $\boldx\leq \boldy$. Notice that any $\boldy\in Y\subseteq E(n,2)$ can be written as the sum of two weight-one words of length $n$. Let us denote by $\bolde_j^n$ the word of length $n$ whose only nonzero coordinate is a $1$ at index $j$. Then every $\boldy\in Y$ can be written $\boldy=\bolde_j^n+\bolde_k^n$. Let $\mathcal{Y}'\subseteq E(2n,2)$ be defined by the union of 3 sets of codeword classes as follows.
  \begin{enumerate}
  \item For each $Y\in \mathcal{Y}$, the codeword class
    \begin{equation} \notag
      \{\bolde_j^{2n}+\bolde_k^{2n} | \bolde_j^n+\bolde_k^n \in Y\}\cup\{\bolde_{j+n}^{2n}+\bolde_{k+n}^{2n} | \bolde_j^n+\bolde_k^n \in Y\}.
    \end{equation}
  \item For each $Y\in \mathcal{Y}$, the codeword class
    \begin{equation} \notag
      \{\bolde_j^{2n}+\bolde_{k+n}^{2n} | \bolde_j^n+\bolde_k^n \in Y\}\cup\{\bolde_{j+n}^{2n}+\bolde_{k}^{2n} | \bolde_j^n+\bolde_k^n \in Y\}.
    \end{equation}
  \item The codeword class
    \begin{equation} \notag
      \{\bolde_j^{2n}+\bolde_{j+n}^{2n} | 1\leq j \leq n\}.
    \end{equation}
  \end{enumerate}
  These codeword classes are trivially disjoint, and each of them covers $E(2n,1)$. Thus, $A(2n,2) \geq A(n,2)+A(n,2)+1 = 2A(n,2)+1$.
\end{IEEEproof}

\begin{proposition}
  \label{prop:aqn1}
  For any $n\geq 1$, $A(n,1)=n$.
\end{proposition}

\begin{IEEEproof}
Partition $E(n,1)$ into $n$ singletons $\{\bolde_j^n\}$ for $1\leq j\leq n$.
\end{IEEEproof}

\begin{proposition}
  \label{prop:a22^n2}
  For any $n \geq 0$, $A(2^n,2)=2^n-1$.
\end{proposition}

\begin{IEEEproof}
  We use $A(2n,2)\geq 2A(n,2)+1$ from Proposition~\ref{prop:a22n2} and the simple bound $A(n,i)\leq n-i+1$, which for $i=2$ becomes $A(n,2)\leq n-1$, and proceed by induction. $A(1,2)=0$. Assuming $A(2^n,2)=2^n-1$, we have $A(2^{n+1},2)\geq 2\cdot (2^n-1)+1=2^{n+1}-1$, and we have $A(2^{n+1},2)\leq 2^{n+1}-1$.
\end{IEEEproof}

\begin{proposition}
  \label{prop:aq2n+12}
  For any $n\geq 1$, $A(2n+1,2)\leq 2n-1$.
\end{proposition}

\begin{IEEEproof}
  This is the bound $A(n,i)\leq\displaystyle\left\lfloor\frac{{n\choose i}}{\left\lceil \frac{{n\choose i-1}}{i} \right\rceil}\right\rfloor$ applied to $i=2$.
\end{IEEEproof}

\begin{proposition} \label{prop:2n+1}
  For any $n\geq 1$, $A(2^n+1,2)=2^n-1$.
\end{proposition}

\begin{IEEEproof}
  $A(2^n+1,2)\geq 2^n-1$ comes from a direct use of Proposition~\ref{prop:aqni_vs_aqn-1} on the results of Propositions~\ref{prop:aqn1} and \ref{prop:a22^n2}. $A(2^n+1,2)\leq 2^n-1$ comes from Proposition~\ref{prop:aq2n+12}.
\end{IEEEproof}

\subsection{Laminar WOM Codes with $n>t$}
\label{sec:w(imi)_disjoint}

The constraint $n=t$ results in relatively low WOM-rates. Lifting this constraint allows to achieve higher WOM-rates. Laminar codes with $n$ slightly larger than $t$ can easily be derived from the codes with $n=t$ above by merging several generations together: taking, as the new set of codeword classes, the union of the sets of codeword classes of two or more consecutive generations.\footnote{$n$ should remain small, because we do not expect to find synchronous codes of WOM-rate higher than nondecodable ones, thus a larger number of cells should be reserved to the nondecodable code.} For instance, the $[4,4:4,3,1,1]_2\sync$ code can be turned into a $[4,3:4,3,2]_2\sync$ code by merging its third and fourth generations together. Instead of having one codeword class at generation $3$ ($\{1110,1101,1011,0111\}$) and one at generation $4$ ($\{1111\}$), now the third generation has two codeword classes: $\{1110,1101,1011,0111\}$ and $\{1111\}$, and there is no fourth generation anymore. Likewise, a $[5,3:5,3,4]_2\sync$ code (of WOM-rate $1.181$) can be derived from the $[5,5:5,3,2,1,1]_2\sync$ code by merging the last three generations together. However, consider the codeword classes of vectors of weight $4$. These were constructed in order to cover every word of weight $3$, while they now only have to cover every word of weight $2$. The optimization also did not allow codeword classes of mixed weights. We can reorganize the set of vectors of weight $3$ or more into a better balanced set of codeword classes. In (\ref{eq:generationsW1}), we give the codeword classes of the third generation of a $[5,3:5,3,6]_2\sync$ code (of WOM-rate $1.298$) obtained by reorganizing the third generation of the $[5,3:5,3,4]_2\sync$ code,
\ifonecolumn
\else
\begin{align}
\label{eq:generationsW1}
&\{01111,11001,10110\},~\{10111,11100,01011\},\nonumber\\ 
&\{11011,01110,10101\},~\{11101,00111,11010\},\nonumber\\ 
&\{11110,10011,01101\},~\{11111\}.
\end{align}
\fi


For comparison, the $4$ codeword classes of the third generation of the $[5,3:5,3,4]_2\sync$ code are
\begin{align}
\label{eq:generationsW1bis}
&\{11100,11010,10101,01011,00111\}~(\mathrm{weight}~3~\mathrm{only}),\nonumber\\
&\{11001,10110,10011,01110,01101\}~(\mathrm{weight}~3~\mathrm{only}),\nonumber\\
&\{11110,11101,11011,10111,01111\}~(\mathrm{weight}~4~\mathrm{only}),\nonumber\\
&\{11111\}~(\mathrm{weight}~5).
\end{align}

Other choices can be made regarding which generations to merge to obtain a $3$-write code from the $[5,5:5,3,2,1,1]_2\sync$ code, but lower WOM-rates are obtained.

\section{A Construction for Synchronous WOM Codes of Higher $t$}
\label{sec:construction}

In this section, we propose a construction to obtain synchronous codes for higher values of $t$ by concatenating $n'$ instances of a synchronous code of length $n$, and using a second synchronous code of length $n'$ to decide, at each generation, which of the $n'$ instances of the first code are going to be modified.
\begin{theorem}
\label{theo:construction}
Let $C$ be a binary $[n,t:M_1,\dots,M_t]_2$ synchronous code of WOM-rate $R$, and $C'$ a binary $[n',t':M_1',\dots,M_{t'}']_2$ synchronous code of WOM-rate $R'$, both not containing the all-zero codeword. Then there exists a binary $[nn',tt':M_1M_1',\dots,M_1M_{t'}',\dots,M_tM_1',\dots,M_tM_{t'}']_2$ synchronous code $C_1$ of WOM-rate $R_1=\frac{t'}{n'}R+\frac{t}{n}R'$.
\end{theorem}

This construction is based on three algorithms.
\begin{enumerate}
\item An algorithm to determine the current generation $i$ of $C_1$ from the state of the $nn'$ memory cells.
\item An encoding algorithm, whose input range depends on $i$.
\item A decoding algorithm.
\end{enumerate}

For $p\in\{1,\dots,t\}$ and $l\in\{1,\dots,t'\}$, we denote by $\calE_p$ and $\calD_p$ the encoding and decoding mappings, respectively, of $C$ at generation $p$, and by $\calE_l'$ and $\calD_l'$ the encoding and decoding mappings, respectively, of $C'$ at generation $l$. We also write $\Image(\calE_0)=\{\mathbf{0}_n\}$ (resp.\ $\Image(\calE'_0)=\{\mathbf{0}_{n'}\}$) to denote the fact that the state of a block prior to any write by $C$ (resp.\ $C'$) is the all-zero codeword of length $n$ (resp.\ $n'$). We then denote by $g$ (resp.\ $g'$) the function that takes a codeword from $C$ (resp.\ $C'$) and returns the unique generation of $C$ (resp.\ $C'$) of which it is a codeword. Formally,
\begin{equation} \notag
  \begin{array}{rrcl}
    g: & \bigcup_{p\in \{0,\dots,t\}} \Image(\calE_p)   & \to     & \{0,\dots,t\}                         \\
       & \boldb                                     & \mapsto & p\text{ s.t. }\boldb\in \Image(\calE_p),  \\
   g': & \bigcup_{l\in \{0,\dots,t'\}} \Image(\calE'_l) & \to     & \{0,\dots,t'\}                        \\
       & \boldb'                                    & \mapsto & l\text{ s.t. }\boldb'\in \Image(\calE'_l).
  \end{array}
\end{equation}
The fact that $C$ and $C'$ are synchronous guarantees that $p$ and $l$ are unique.


\begin{algorithm}
\caption{Algorithm to Recover the Current Generation} \label{alg:gen}
\begin{algorithmic}[1]
\State Input: $\mathbf{b}_1,\dots,\mathbf{b}_{n'}$
\State Output: $p,l,i$, and $\mathbf{b}'$
\State $p\gets 0$
\For{$k\gets 1 \algorithmicto n'$}
  \State $p_k\gets g(\boldb_k)$
  \If{$p_k>p$}
    \State $p=p_k$
  \EndIf
\EndFor
\For{$k\gets 1 \algorithmicto n'$}
  \State $b'_k\gets p_k+1-p$ \Comment{Should always be $0$ or $1$}
\EndFor
\If{$\mathbf{b}' = \mathbf{1}_{n'}$ or $g'(\boldb') = t'$}
  \State $l\gets 0$
  \State $p\gets p+1$
  \State $\boldb'\gets \mathbf{0}_{n'}$
\Else
\State $l\gets g'(\boldb')$
\EndIf
\State $i\gets (p-1)t'+l$
\end{algorithmic}
\end{algorithm}

The key idea is that the $nn'$ wits of $C_1$ are divided into $n'$ blocks of $n$ wits denoted by $\boldb_k$ for $k\in\{1,\dots,n'\}$, and the $tt'$ generations are divided into $t$ stages of $t'$ generations. For $p\in\{1,\dots,t\}$ and $l\in\{1,\dots,t'\}$, generation $i=(p-1)t'+l$ of $C_1$ is the $l$-th generation of the $p$-th stage. At this point, we guarantee that each of the $n'$ blocks of $n$ wits contains a codeword $\boldb_k\in \Image(\calE_{p-1}) \cup \Image(\calE_p)$. We call $\boldb'=(b'_1,\dots,b'_{n'})\in \Image(\calE'_l)$ the binary vector of length $n'$ with entries $b'_k=g(\boldb_k)-p+1$, $k\in\{1,\dots,n'\}$. Then, Algorithm \ref{alg:gen} can take a codeword of $C_1$, and use functions $g$ and $g'$ to determine the current generation $i$.

Both the encoder and the decoder first use this algorithm to determine the current generation $i$ (actually, they use $p$ and $l$). They also use the value of $\boldb'$. Algorithm~\ref{alg:enc}, described below,  is the encoding algorithm, which takes a message $m_1\in\{1,\dots,M_pM'_{l+1}\}$ and encodes it. This message is decomposed into a message $m\in\{1,\dots,M_p\}$ and a message $m'\in\{1,\dots,M'_{l+1}\}$. We then compute the new $\boldb'$ as $\calE'_{l+1}(m',\boldb')$ and compare the positions at which it differs from the old one. These positions are the indices $k\in\{1,\dots,n'\}$ of the blocks that will be written (hence switching from generation $p-1$ to generation $p$). The only requirement on how these blocks will be written is that after this encoding stage, the modulo $M_p$ sum (in $\{1,\dots,M_p\}$) of the $\calD_p(\boldb_k)$ for $\boldb_k\in \Image(\calE_p)$ is $m$. Algorithm \ref{alg:enc} shows a simple way to achieve this.

\begin{algorithm}
\caption{Encoding Algorithm} \label{alg:enc}
\begin{algorithmic}[1]
\State Input: $\mathbf{b}_1,\dots,\mathbf{b}_{n'}$, and $\mathbf{b}'$, $m_1$, $p$, and $l$
\State Output: $\mathbf{b}_1,\dots,\mathbf{b}_{n'}$
\State $m\gets 1+\left \lfloor (m_1-1)/M'_{l+1} \right\rfloor$
\State $m'\gets 1+((m_1-1)\ \mathrm{mod}\ M'_{l+1})$
\State $\hat{\boldb}'\gets \calE'_{l+1}(m',\boldb')$
\For{$k\gets 1 \algorithmicto n'$}
  \If{$b'_k=1$}
    \State $m\gets m-\calD_p(\boldb_k)$
  \ElsIf{$b'_k=0\ \wedge \ \hat{b}'_k=1$}
    \State $k_0\gets k$
  \EndIf
\EndFor
\State $m\gets 1+((m-1)\ \mathrm{mod}\ M_p)$
\For{$k\gets 1 \algorithmicto n'$}
  \If{$\hat{b}'_k=1\ \wedge \ b'_k=0\ \wedge \ k\neq k_0$}
    \State $\boldb_k\gets \calE_p(M_p,\boldb_k)$
  \EndIf
\EndFor
\State $\boldb_{k_0}\gets \calE_p(m,\boldb_{k_0})$
\end{algorithmic}
\end{algorithm}

The messages $m$ and $m'$ can be decoded by decoding $\boldb'$ with the decoder of $C'$, and then decoding every block $\boldb_k$ for $\boldb_k\in \Image(\calE_p)$ with the decoder of $C$, and finally taking the modulo $M_p$ sum (in $\{1,\dots,M_p\}$) of the decoded messages. The original message $m_1$ is then $m_1=(m-1)M'_l+m'$. See Algorithm \ref{alg:dec} for details.

\begin{algorithm}
\caption{Decoding Algorithm} \label{alg:dec}
\begin{algorithmic}[1]
\State Input: $\mathbf{b}_1,\dots,\mathbf{b}_{n'}$, and $\mathbf{b}'$, $p$, and $l$
\State Output: $m$, $m'$, and $m_1$
\State $m'\gets \calD'_l(\boldb')$
\State $m\gets 0$
\For{$k\gets 1 \algorithmicto n'$}
  \If{$b'_k=1$}
    \State $m\gets m+\calD_p(\boldb_k)$
  \EndIf
\EndFor
\State $m\gets 1+((m-1)\ \mathrm{mod}\ M_p)$
\State $m_1\gets (m-1)M'_l+m'$
\end{algorithmic}
\end{algorithm}

Let us now establish the WOM-rate $R_1$ of $C_1$.
\ifonecolumn
\else
\begin{equation} \notag
  \begin{split}
    R_1 & =\frac{\sum_{p=1}^t\sum_{l=1}^{t'}\log_2(M_pM_l')}{nn'}                                                        \\
        & =\frac{1}{nn'}\left(\log_2\left(\prod_{p=1}^tM_p^{t'}\right)+\log_2\left(\prod_{l=1}^{t'}(M_l')^t\right)\right)  \\
        & =\frac{1}{nn'}\left(t'\cdot nR+t\cdot n'R'\right)  =\frac{t'}{n'}R+\frac{t}{n}R'.
  \end{split}
\end{equation}
\fi

\begin{example}
\label{sec:example}
Let $C$ be the $[4,3:4,3,2]_2\sync$ code defined by
\ifonecolumn
\else
   \begin{equation*} \notag
     \hspace{-.02\textwidth}
    \begin{array}{r|cccc}
                   & 1                       & 2             & 3             & 4           \\ \hline
      \calD_1^{-1} & \{0001\}                & \{0010\}      & \{0100\}      & \{1000\}    \\
      \calD_2^{-1} & \{1100,0011\}           & \{1010,0101\} & \{1001,0110\} & -           \\
      \calD_3^{-1} & \begin{array}{c}
                       \{0111,1011,\\
                       1101,1110\} 
                      \end{array}            & \{1111\}      & -             & -           \\
    \end{array}
  \end{equation*}
\fi
and $C'$ the $[2,2:2,1]_2\sync$ code defined by
\ifonecolumn
\else
  \begin{equation*} \notag
    \begin{array}{r|cccc}
                      & 1       & 2       \\ \hline
      (\calD_1')^{-1} & \{01\}  & \{10\}  \\
      (\calD_2')^{-1} & \{11\}  &  -      \\
    \end{array}.
  \end{equation*}
\fi

The code $C_1$ obtained with the construction is a $[8,6:8,4,6,3,4,2]_2\sync$ code. Consider that the eight cells are in state $(\boldb_1,\boldb_2)=(1100,0010)$. Let us first consider the decoding of the message following Algorithms~\ref{alg:gen} and \ref{alg:dec}. The generation in $C$ of the first block $\boldb_1$ is $2$, and that of the second block $\boldb_2$ is $1$, thus $p=2$ (the highest of the two) and $\boldb'=(10)$. The fact that $C'$ is synchronous guarantees that only one encoding function  of $C'$ has $\boldb'$ in its range: here, it is the encoding function for $l=1$. Thus, we are at the first generation ($l=1$) of the second stage ($p=2$), so the overall generation is $i=(p-1)t'+l=(2-1)\times 2+1=3$. The flow of Algorithm~\ref{alg:gen} is illustrated in Fig.~\ref{fig:ex_alg_gen}. 

For the decoding part, we have $m'=\calD_1'(10)=2$ and $m$ as the modulo $M_p$ sum (in $\{1,\dots,M_p\}$) of $\calD_p(\boldb_k)$ for all indices $k$ of a block at generation $p$ of $C$. Here, there is only one block at generation $p=2$ for $C$: block $\boldb_1=(1100)$, therefore $m=(\calD_2(\boldb_1)-1) \pmod{3}+1=1$. The original message pair was therefore $(1,2)$. This can be mapped to $m_1\in\{1,\dots,M_p M_l'\}$ by $m_1=(m-1)M_l'+m'$,  which gives $m_1=0\times 2 + 2 = 2$. The flow of Algorithm~\ref{alg:dec} is illustrated in Fig.~\ref{fig:ex_alg_dec}.
\begin{figure*}[htbp]
\centering

\begin{tikzpicture}[font=\footnotesize]
  
  \tikzstyle{tabbox}=[minimum height=.5cm,minimum width=2cm,inner sep=0]

  
  \node[tabbox] (pb1) at ( 0, 1  ) {};
  \node[tabbox] (pb2) at ( 2, 1  ) {};

  \node at (-0.75,1) {$1$};\node at (1.25,1) {$0$};
  \node at (-0.25,1) {$1$};\node at (1.75,1) {$0$};
  \node at ( 0.25,1) {$0$};\node at (2.25,1) {$1$};
  \node at ( 0.75,1) {$0$};\node at (2.75,1) {$0$};

  \draw (-1, 1.25) -- ( 3, 1.25);
  \draw (-1, 0.75) -- ( 3, 0.75);
  
  \draw (-1, 1.25) -- (-1, 0.75);
  \draw[dotted] ( 1, 1.25) -- ( 1, 0.75);
  \draw ( 3, 1.25) -- ( 3, 0.75);


  \node[tabbox,\ccolora] (b1) at ( 0, 0  ) {$\boldb_1=1100$};
  \node[tabbox,\ccolora] (b2) at ( 2, 0  ) {$\boldb_2=0010$};
  
  \draw (-1, 0.25) -- ( 3, 0.25);
  \draw (-1,-0.25) -- ( 3,-0.25);
  
  \draw (-1, 0.25) -- (-1,-0.25);
  \draw ( 1, 0.25) -- ( 1,-0.25);
  \draw ( 3, 0.25) -- ( 3,-0.25);

  
  \draw[->] 
  (pb1) edge (b1)
  (pb2) edge (b2);


  \node[tabbox,\ccolora] (g1) at ( 0,-1) {$g(\boldb_1)=2$};
  \node[tabbox,\ccolora] (g2) at ( 2,-1) {$g(\boldb_2)=1$};
  
  \draw (-1,-1.25) -- ( 3,-1.25);
  \draw (-1,-0.75) -- ( 3,-0.75);
  
  \draw (-1,-1.25) -- (-1,-0.75);
  \draw ( 1,-1.25) -- ( 1,-0.75);
  \draw ( 3,-1.25) -- ( 3,-0.75);

  \node[\ccolora,inner sep=3] (p) at (6.2,-1) {$p=2$};
  \draw[->] (g2) edge (p);
  
  
  \draw[->] 
  (b1) edge node[right,\ccolora] {$g()$} (g1)
  (b2) edge node[right,\ccolora] {$g()$} (g2);
  
  
  \node[tabbox] (c1) at ( 0,-2) {$1$};
  \node[tabbox] (c2) at ( 2,-2) {$0$};
  
  \draw (-1,-1.75) -- ( 3,-1.75);
  \draw (-1,-2.25) -- ( 3,-2.25);
  
  \draw (-1,-1.75) -- (-1,-2.25);
  \draw ( 1,-1.75) -- ( 1,-2.25);
  \draw ( 3,-1.75) -- ( 3,-2.25);

  \node[\ccolorb,inner sep=3] (cb) at (4.2,-2) {$\boldb'=10$};
  \draw[->] (c2) edge (cb);
  
  
  \draw[->] 
  (g1) edge (c1)
  (g2) edge (c2);
  
  
  \node[\ccolorb] (gb) at (4.2,-3) {$l=g'(\boldb')=1$};
  \draw[->] (cb) edge node[right,\ccolorb] {$g'()$} (gb);
  
  
  \node (i) at (5.2,-4) {$i=\tcola{(p-1)}\tcolb{t'}+\tcolb{l}=3$};
  \draw[double] (3.7 ,-4.25) -- (6.7 ,-4.25) -- (6.7 ,-3.75) -- (3.7 ,-3.75) -- cycle;
  \node[tabbox] (pq) at (4.2,-4) {};
  \node[tabbox] (pp) at (6.2,-4) {};

  \draw[->]
  (p)  edge (pp)
  (gb) edge (pq);
  

  \draw[decorate,decoration=brace] (-1.25,-0.2) -- node[left] {Input}     (-1.25, 1.2);
  \draw[decorate,decoration=brace] (-1.25,-1.2) -- node[left] {Lines 4-7} (-1.25,-0.3);
  \draw[decorate,decoration=brace] (-1.25,-2.2) -- node[left] {Lines 8,9} (-1.25,-1.3);
  \draw[decorate,decoration=brace] (-1.25,-3.2) -- node[left] {Lines 10-15}   (-1.25,-2.3);
  \draw[decorate,decoration=brace] (-1.25,-4.2) -- node[left] {Line 16}   (-1.25,-3.3);

\end{tikzpicture}

\caption{Example of a run of Algorithm~\ref{alg:gen}.}
\label{fig:ex_alg_gen}
\end{figure*}
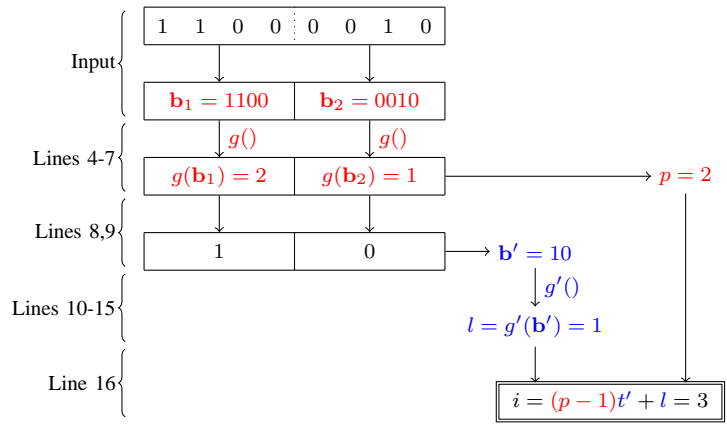
\begin{figure*}[htbp]
\centering

\begin{tikzpicture}[font=\footnotesize]
  
  \tikzstyle{tabbox}=[minimum height=.5cm,minimum width=2cm,inner sep=0]
 

  \node[tabbox,\ccolora           ] (b1) at ( 0, 0) {$\boldb_1=1100$};
  \node[tabbox,\ccolora,opacity=.7] (b2) at ( 2, 0) {$\boldb_2=0010$};
  
  \draw (-1, 0.25) -- ( 3, 0.25);
  \draw (-1,-0.25) -- ( 3,-0.25);
  
  \draw (-1, 0.25) -- (-1,-0.25);
  \draw ( 1, 0.25) -- ( 1,-0.25);
  \draw ( 3, 0.25) -- ( 3,-0.25);

  \node[tabbox,\ccolorb] (bp) at (-2.75, 0) {$\boldb'=10$};

  \draw (-4, 0.25) -- (-1.5, 0.25) -- (-1.5,-0.25) -- (-4,-0.25) -- cycle;

  \draw[decorate,decoration=brace,\ccolora] (-0.95,0.45) -- node[above,\ccolora] {gen. $p=2$}   ( 0.95,0.45);
  \draw[decorate,decoration=brace,\ccolora] ( 1.05,0.45) -- node[above,\ccolora] {gen. $p-1=1$} ( 2.95,0.45);

  \draw[decorate,decoration=brace,\ccolorb] (-3.95,0.45) -- node[above,\ccolorb] {gen. $l=1$}   (-1.55,0.45);


  \node[tabbox,\ccolorb] (dp) at (-2.75,-1) {$m'=\calD'_1(\boldb')=2$};

  \draw (-4,-0.75) -- (-4,-1.25) -- (-1.5,-1.25) -- (-1.5,-0.75) -- cycle;
  

  \draw[->]
  (bp) edge node[right,\ccolorb] {$\calD'_l()$} (dp);


  \node[tabbox,\ccolora]            (d1) at ( 0,-2) {$\calD_2(\boldb_1)=1$};
  \node[tabbox,\ccolora,opacity=.7] (d2) at ( 2,-2) {ignored};
  
  
  
  \draw[->]            (b1) edge node[pos=.82,right,\ccolora]            {$\calD_p()$}  (d1);
  \draw[->,opacity=.7] (b2) edge node[pos=.82,right,\ccolora,opacity=.7] {not gen. $p$} (d2);
  
  
  \draw[decorate,decoration=brace,\ccolora] (2.95,-2.45) -- node[tabbox,below=3pt,\ccolora] (sum) {$\sum_k\calD_2(\boldb_k)=1$} (-0.95,-2.45);


  \node[tabbox,\ccolora] (d) at (1,-4) {$m=1$};

  
  \draw[->] (sum) edge node[right,\ccolora] {proj. to $\{1,\dots,M_p\}$} (d);


  \node (m1) at (-0.875,-5) {$m_1=\tcola{(m-1)}\tcolb{M_l'}+\tcolb{m'}=2$};
  \draw[double] (-3.75,-5.25) -- (2,-5.25) -- (2,-4.75) -- (-3.75,-4.75) -- cycle;
  \node[tabbox] (pdp) at (-2.75,-5) {};
  \node[tabbox] (pd)  at ( 1   ,-5) {};


  \draw[->]
  (dp) edge (pdp)
  (d)  edge (pd);


  \draw[dotted] (-1.25,0.9) -- (-1.25,-4.6);

  \draw[decorate,decoration=brace] (-4.25,-0.2) -- node[left] {Input}     (-4.25, 0.75);
  \draw[decorate,decoration=brace] (-4.25,-1.2) -- node[left] {Line 3}    (-4.25,-0.3 );
  \draw[decorate,decoration=brace] (-4.25,-4.2) -- node[left] {Lines 4-8} (-4.25,-1.3 );
  \draw[decorate,decoration=brace] (-4.25,-5.2) -- node[left] {Line 9}    (-4.25,-4.3 );

\end{tikzpicture}

\caption{Example of a run of Algorithm~\ref{alg:dec}.}
\label{fig:ex_alg_dec}
\end{figure*}
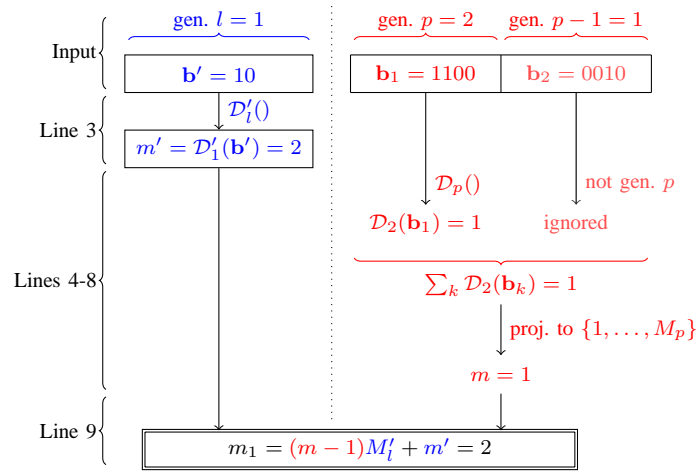

For the encoding part, let us now encode a new message $m_1=2\in\{1,2,3\}$ for generation $4$  following Algorithm~\ref{alg:enc}. Our new $m$ and $m'$ are $2$ and $1$, respectively, so that $(m-1)M_{l+1}'+m'=(2-1)\times 1+1=2$. $\boldb'=(10)$ will become $\boldb'=(11)$ because $\calE_2'(1,10)=(11)$. Therefore, the second block is going to be written (because the second wit of $\boldb'$ changes). We first decode all the blocks already at generation $p=2$: here, we only have one block at generation $p=2$, and $\calD_2(\boldb_1)=\calD_2(1100)=1$. We therefore encode in the second block $\boldb_2$ a message $m_0 = (m-\calD_2(\boldb_1)-1) \pmod {M_p}+1$, 
where $M_p=M_2=3$ and $m=2$. Thus, $m_0=1$ and $\boldb_2$ is replaced by $\calE_p(1,0010)=(0011)$. The state of the cells is $(1100,0011)$ after this encoding phase. The flow of Algorithm~\ref{alg:enc} is illustrated in Fig.~\ref{fig:ex_alg_enc}.
\begin{figure*}[htbp]
\centering

\begin{tikzpicture}[font=\footnotesize]
  
  \tikzstyle{tabbox}=[minimum height=.5cm,minimum width=2cm,inner sep=0]
  

  \node[tabbox,draw,minimum width=1.5cm] (m1) at (-1.25,1) {$m_1=2$};

  \node[tabbox,\ccolorb,draw=black,minimum width=1.5cm] (mp) at (-2.25,0) {$m'=1$};
  \node[tabbox,\ccolora,draw=black,minimum width=1.5cm] (m)  at (-0.25,0) {$m=2$};
  
  \draw[->]
  (m1) edge (mp)
  (m1) edge (m);

  \draw[dotted] (-1.25,0.65) -- (-1.25,-4.3);


  \begin{scope}[xshift=2cm]
    \node[tabbox,\ccolora           ] (b1) at ( 0, 1) {$\boldb_1=1100$};
    \node[tabbox,\ccolora,opacity=.7] (b2) at ( 2, 1) {$\boldb_2=0010$};
  
    \draw (-1,1.25) -- ( 3,1.25);
    \draw (-1,0.75) -- ( 3,0.75);
  
    \draw (-1,1.25) -- (-1,0.75);
    \draw ( 1,1.25) -- ( 1,0.75);
    \draw ( 3,1.25) -- ( 3,0.75);

    \draw[decorate,decoration=brace,\ccolora] (-0.95,1.45) -- node[above,\ccolora] {gen. $p=2$}   ( 0.95,1.45);
    \draw[decorate,decoration=brace,\ccolora] ( 1.05,1.45) -- node[above,\ccolora] {gen. $p-1=1$} ( 2.95,1.45);

    \node[tabbox,\ccolora]            (d1) at ( 0,-3.5) {$\calD_2(\boldb_1)=1$};
    \node[tabbox,\ccolora,opacity=.7] (d2) at ( 2,-3.5) {ignored};
    
    \draw[->]            (b1) edge node[pos=.96,right,\ccolora]            {$\calD_p()$}  (d1);
    \draw[->,opacity=.7] (b2) edge node[pos=.96,right,\ccolora,opacity=.7] {not gen. $p$} (d2);

    \draw[decorate,decoration=brace,\ccolora] (2.95,-3.95) -- node[tabbox,below=3pt,\ccolora] (sum) {$\tilde{m}_0=\sum_k\calD_2(\boldb_k)=1$} (-0.95,-3.95);

  \end{scope}

  
  \begin{scope}[xshift=-2cm]

    \node[tabbox,\ccolorb] (bp) at (-2.75, 1) {$\boldb'=10$};

\ifonecolumn
\else
    \draw (-4.15, 1.25) -- (-1.35, 1.25) -- (-1.35,0.75) -- (-4.15,0.75) -- cycle;
\fi    
\ifonecolumn
\else
    \draw[decorate,decoration=brace,\ccolorb] (-3.95,1.45) -- node[above,\ccolorb] {gen. $l=1$}   (-1.55,1.45);
\fi

    \node[tabbox,\ccolorb] (ep) at (-2.75,-1) {$\hat{\boldb}'=\calE'_2(m',\boldb')=11$};
    
\ifonecolumn
\else
    \draw (-4.15,-0.75) -- (-4.15,-1.25) -- (-1.35,-1.25) -- (-1.35,-0.75) -- cycle;
\fi
    
    \draw[->] (bp) edge node[pos=.82,right,\ccolorb] (epl) {$\calE'_{l+1}()$} (ep);
    \draw[->] (mp) .. controls +(down:.45cm) and +(right:.9cm) .. (epl);

    \node[tabbox,\ccolorb] (k) at (-2.75,-2.5) {$k_0=2$};
    

    \draw[->] (ep) edge node[right=-0.2cm,\ccolorb] {\begin{tabular}{l}$\boldb'=10$\\ $\hat{\boldb}'=11$\end{tabular}} (k);
\ifonecolumn
\else
    \draw[dashed] (-1.70,-1.75) ellipse (0.11cm and 0.4cm);
\fi

  \end{scope}

  \node[draw,tabbox] (newb2) at (-1.25,-6) {\ \ $\tcola{\boldb_2}\gets\tcola{\calE_2(m\ominus \tilde{m}_0,\boldb_2)}=\tcola{0011}$\ \ };
\ifonecolumn
\else
  \node at (4.5,-6) {(where $\ominus$ is the modulo $M_p$ subtraction in $\{1,\dots,M_p\}$)};
\fi
  \begin{scope}[xshift=-2.25cm,yshift=-8cm]
    \node at (-0.75,1) {$1$};\node at (1.25,1) {$0$};
    \node at (-0.25,1) {$1$};\node at (1.75,1) {$0$};
    \node at ( 0.25,1) {$0$};\node at (2.25,1) {$1$};
    \node at ( 0.75,1) {$0$};\node at (2.75,1) {$1$};

    \node[tabbox] (new) at (1,1) {};
    
    \draw[double] (-1, 1.25) -- ( 3, 1.25) -- ( 3, 0.75) -- (-1, 0.75) -- cycle;
    \draw[dotted] ( 1, 1.25) -- ( 1, 0.75);
  \end{scope}
  
  \draw[->]
  (k)     edge (newb2)
  (m)     edge (newb2)
  (sum)   edge (newb2)
  (newb2) edge (new);

  \draw[decorate,decoration=brace] (-6.35, 0.8) -- node[left] {Input}       (-6.35, 2.1);
  \draw[decorate,decoration=brace] (-6.35,-0.2) -- node[left] {Lines 3,4}   (-6.35, 0.7);
  \draw[decorate,decoration=brace] (-6.35,-1.2) -- node[left] {Line 5}      (-6.35,-0.3);
  \draw[decorate,decoration=brace] (-6.35,-2.7) -- node[left] {Lines 9,10}   (-6.35,-1.3);
  \draw[decorate,decoration=brace] (-6.35,-4.7) -- node[left] {Lines 7,8,11} (-6.35,-2.8);
  \draw[decorate,decoration=brace] (-6.35,-6.2) -- node[left] {Lines 12-15} (-6.35,-4.8);
  \draw[decorate,decoration=brace] (-6.35,-7.2) -- node[left] {Output}      (-6.35,-6.3);

\end{tikzpicture}

\caption{Example of a run of Algorithm~\ref{alg:enc}.}
\label{fig:ex_alg_enc}
\end{figure*}
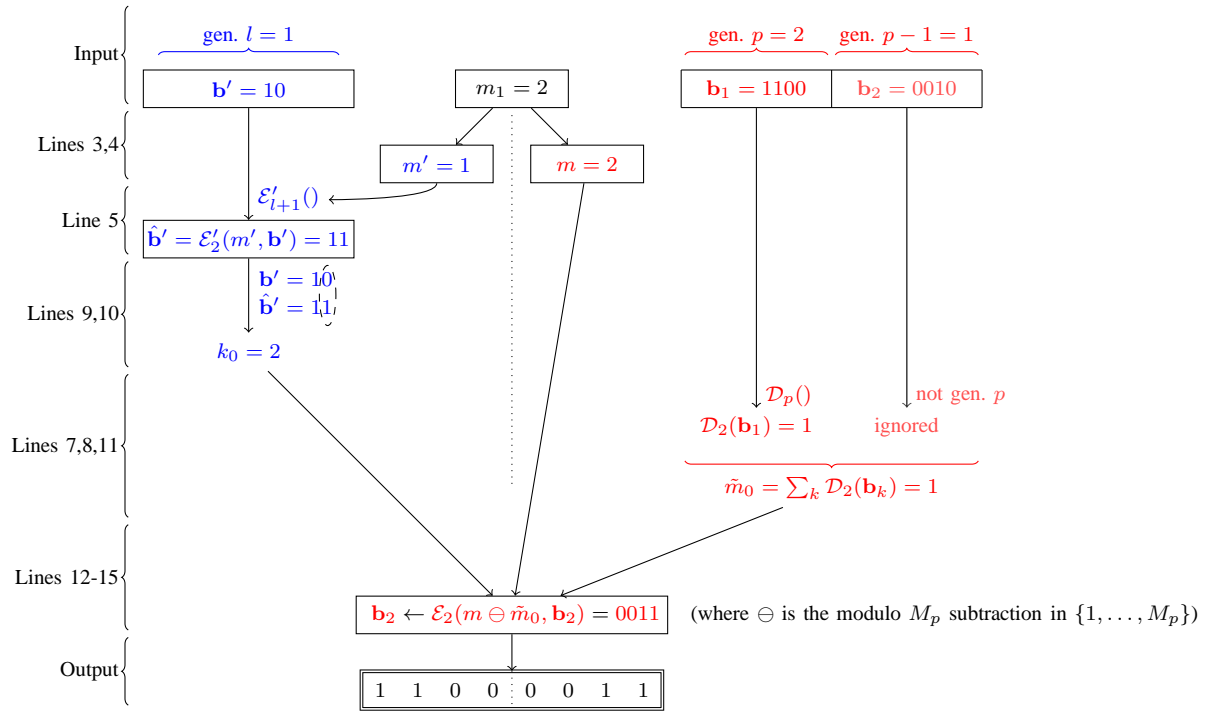

\end{example}

We remark that the construction above requires that code $C$ does not contain the all-zero codeword. In that case, if the all-zero codeword of $C$ is written in a block, the generation of $C_1$ would be improperly identified and the component $m'$ of the message could not be written/decoded. The construction also requires $C'$ to not contain the all-zero codeword, in which case the component $m$ of the message could not be written/decoded when the all-zero codeword is chosen for $C'$.

As a final remark, note that the construction above resembles a tensor-product code construction, but with some important differences. For instance, it is required that the different blocks contain codewords from $C$ of neighboring generations.

\subsection{Results}

Let us denote by $F(C,C')$ the code obtained by applying the construction of Theorem~\ref{theo:construction} to $C$ and $C'$. We can iterate the above construction by choosing $C$ and $C'$, and then defining $C_0=C$ and $C_m=F(C_{m-1},C')$ for all $m>0$. This generates codes with even higher values of $t$, which have to be compared with a construction of synchronous codes from \cite{RivSha82} (where $n=t$ is any power of two and the WOM-rate is $\log_2(t)/2$). Notice that the two constructions happen to match when we take as $C=C'$ the trivial $[2,2:2,1]_2\sync$ code.

First, we restrict ourselves to codes with $n=t$ (which are easier to compare) and we fix $C'=C$. The WOM-rate of the $t_m$-write code $C_m$ after $m$ iterations of the construction is
\begin{equation} \notag
  R(C_m)=mR(C)=\log_t(t_m)R(C)=\frac{R(C)}{\log_2(t)}\log_2(t_m).
\end{equation}
Therefore, for codes with $n=t$, the higher $\frac{R(C)}{\log_2(t)}$ is, the better this iterated construction works. The code that maximizes this ratio among those found by our computer search is the one with $n=t=2$ (with $\frac{R(C)}{\log_2(t)}=\frac{1}{2}$), making the codes from \cite{RivSha82} the best in terms of asymptotic WOM-rate until codes for higher values of $n=t$ are found. For instance, Table~\ref{table:bound_2} suggests that a $[8,8:8,7,5,5,2,2,1,1]_2\sync$ code could exist, with a ratio of $0.519$ (and even better synchronous codes could exist even for $n=t=8$, if we remove the added constraints from Section~\ref{sec:w(imi)=i}). However, when $t$ is not a power of two, our construction can yield codes where  $t$ has either $2$, $3$, or $5$ as a divisor, but no other prime divisors, i.e., the number of writes is of the form $2^a3^b5^c$. This is achieved by mixing different elementary codes $C'$ with $2$, $3$, and $5$ generations, instead of always using the $t=2$ code. This is a much denser coverage of the potential values of $t$. Furthermore, if we consider codes with $n$ slightly greater than $t$, we can reach higher WOM-rates at equal values of $t$. Consider, for instance, the code $F(C,C')$ with $C$ the $[4,3:4,3,2]_2\sync$ code and $C'$ the $[2,2:2,1]_2\sync$ code. The construction then yields a $[8,6:8,4,6,3,4,2]_2\sync$ code of WOM-rate $1.521$ (larger than $\log_2(t)/2$ both for $t=6$ and $t=8$). This is the example code of Example~\ref{sec:example}.

\section{Fixed-Rate WOM Codes}
\label{sec:fixed_rate}

In Sections~\ref{sec:w(imi)=i} and~\ref{sec:construction} we did not impose any constraint on the values $\{M_i\}$. Therefore, the obtained codes are in general unrestricted-rate codes, i.e., the codes store in general a different number of messages at different generations. Appending these codes to a nondecodable code to make it decodable will clearly result into an unrestricted-rate code.

In this section, given a fixed-rate nondecodable code, we consider the problem of efficiently generating a fixed-rate decodable code. Note that the standard method of appending $t_{\rm nd}-1$ cells to a $t_{\rm nd}$-write nondecodable code that only store the current generation results in a fixed-rate code, since it does not change the values of $\{M_i\}$. However, we can also improve the WOM-rate of the overall code, by appending a short synchronous code as in the previous sections, with the additional constraint that the synchronous code must also be fixed-rate. We are therefore interested in finding short synchronous fixed-rate codes.

The main result of this section is that the construction of Section~\ref{sec:construction} yields a fixed-rate code when applied to two fixed-rate codes. To find fixed-rate synchronous codes for many values of $t$, one therefore only has to find a few such codes for small values of $t$. In the following, we propose two such codes.
\begin{itemize}
\item A $[3,2:2,2]_2\sync$ code of WOM-rate $2/3$ given by
\ifonecolumn
\else
   \begin{equation*} \notag
     \hspace{-.02\textwidth}
    \begin{array}{r|cccc}
                   & 1                       & 2            \\ \hline
      \calD_1^{-1} & \{001\}                 & \{010\}      \\
      \calD_2^{-1} & \{110,101\}             & \{011\}      \\
    \end{array}.
  \end{equation*}
\fi
\item A $[5,3:4,4,4]_2\sync$ code of WOM-rate $1.2$ where the classes are:
  \begin{itemize}
    \item At generation $1$: $\{00001\}$, $\{00010\}$, $\{00100\}$, and $\{01000\}$.
    \item At generation $2$: $\{11000,10100,10010,10001\}$, $\{01100,00011\}$, $\{01010,00101\}$, and $\{01001,00110\}$.
    \item At generation $3$: the same codeword classes as in (\ref{eq:generationsW1bis}).
  \end{itemize}
\end{itemize}

We remark that fixed-rate codes have not only lower WOM-rate than unrestricted-rate codes, but when we add the constraint that the codes must be synchronous and with $n=t-1$, this gets even worse as the last generation of a synchronous code with $n=t-1$ will always have size $1$, forcing the size of every generation to be $1$ for a fixed-rate code, and making its WOM-rate $0$. This explains why the two codes that we give have $n$ larger than $t-1$.

\section{Extension to $q$-ary WOM Codes}
\label{sec:qary}

The proposed method of Section~\ref{sec:MainIdea} for making a nondecodable code decodable in the binary case can be extended to the problem of making nondecodable $q$-ary codes decodable for $q>2$. The number of additional cells required to make a $q$-ary $\tnd$-write nondecodable code decodable is 
$\left\lceil \frac{\tnd-1}{q-1}\right\rceil$. 
Indeed, during each of the last $\tnd-1$ generations, the sum of the values in the additional cells is increased by at least $1$, and this sum is at most $q-1$ times the number of additional cells. We consider the problem of building synchronous $q$-ary $(\tnd-1)$-write codes with length $\left\lceil \frac{\tnd-1}{q-1}\right\rceil$ (or slightly above) which do not contain the all-zero codeword, since, as in the binary case, we can later add an extra generation  containing only the all-zero codeword, turning the code into a $\tnd$-write code of length  $\left\lceil \frac{\tnd-1}{q-1}\right\rceil$. If $\tnd \leq q$, then only one additional cell is required. This case applies to the codes in \cite{nonbinarywomcodes}, for instance, with $q=8$ and $\tnd=2,3,4,5,6,7$, or $q=4$ and $\tnd=2,3,4$. Then, the WOM-rate of a code is determined entirely by the assignment of the $q$ possible values of the cell to its generations. For instance, if $q=5$ and $\tnd=3$, we can choose $\Image(\calE_1)=\{0,1\}$, $\Image(\calE_2)=\{2,3\}$, and $\Image(\calE_3)=\{4\}$. The WOM-rate of the resulting code would therefore be $\log_2(2\times 2\times 1)$. Maximizing the WOM-rate of the code is equivalent to maximizing the product $\prod_{i=1}^{\tnd}M_i$ where the only constraints on the $M_i$'s are that they are integers from $\{1,\dots,q\}$ and that $\sum_{i=1}^{\tnd}M_i\leq q$. Maximizing a product of integers given their sum is achieved by choosing them as close to each other as possible, here by picking $M_i\in\{\lfloor q/\tnd\rfloor,\lceil q/\tnd \rceil\}$ for all $i$. Let us consider the two extreme regimes. If $\tnd=q/2$ (resp.\ $\tnd>q/2$), we pick $M_i\in\{2,2\}$ (resp.\  $M_i\in\{1,2\}$) and the resulting WOM-rate is $\log_2(2^{q-\tnd})=q-\tnd$ (resp.\ $\log_2(2^{q-\tnd})=q-\tnd$), while if $\tnd$ is small compared to $q$, the optimal WOM-rate can be closely approximated by $\log_2\left(\prod_{i=1}^{\tnd}q/\tnd\right)=\tnd\log_2(q/\tnd)$.

If $\tnd > q$, then several additional cells are required. Using a computer search, we can find a few very short synchronous codes for $q>2$ under the same constraints as the codes from Section~\ref{sec:w(imi)=i} (laminar, with $n=\left\lceil \frac{t}{q-1}\right\rceil$, and where generation $i$ is built assuming that all codewords of weight (or $\ell_1$-norm) $i-1$ are used by generation $i-1$). Furthermore, in analogy with the binary case, we make the following important definition. 
\begin{definition}
Let $E_q(n,i)$ be the set of $q$-ary vectors of length $n$ and weight  $i$, and $A_q(n,i)$ the maximum size of a partition $\mathcal{Y}$ of $E_q(n,i)$ so that
\begin{equation} \notag
  \forall Y\in\mathcal{Y},\,\forall \boldx\in E_q(n,i-1),\,\exists \boldy\in Y:\,\boldx\leq \boldy.
\end{equation}
\end{definition}

As in the binary case, we would like to compute $A_q(n,i)$ for different values of $n$ and $i$.
\ifonecolumn
\else
Tables \ref{tab:sec2q3} and \ref{tab:sec2q4} show the results 
\fi
of such a search for $q=3,4$ and small values of $n$. As an example, a $[2,6:2,2,2,1,1,1]_4\sync$ code  of WOM-rate $3/2$ 
\ifonecolumn
\else
(which corresponds to the second row of Table~\ref{tab:sec2q4}) given by
\fi
\ifonecolumn
\else
   \begin{equation*} \notag
     \hspace{-.02\textwidth}
    \begin{array}{r|cccc}
                   & 1                       & 2            \\ \hline
      \calD_1^{-1} & \{01\}                 & \{10\}      \\
      \calD_2^{-1} & \{11\}             & \{20,02\}      \\
      \calD_3^{-1} & \{21,03\}             & \{12,30\}      \\
      \calD_4^{-1} & \{13,31,22\}             &  -     \\
      \calD_5^{-1} & \{23,32\}             &   -    \\
      \calD_6^{-1} & \{33\}             &     -  \\
    \end{array}
  \end{equation*}
\fi
was found.

\ifonecolumn
\else
\begin{table}[!t]
\footnotesize
\centering
\caption{$A_q(n,i)$ in the ternary case ($q=3$). The values are constructive (i.e., they correspond to actual codes found by an exhaustive search). Values in italics can also be taken from Proposition~\ref{prop:n1q}.}
\begin{tabular}{r|rrrrrrrr}
~~~$i$&1&2&3&4&5&6&7&8\\
$n$~~~& & & & & & & & \\\hline
1   &\em 1&1& & & & & & \\
2   &\em 2&2&1&1& & & & \\
3   &\em 3&3&2&1&1&1& & \\
4   &\em 4&4&3&3&1&1&1&1\\
5   &\em 5&5&-&-&-&-&-&-
\end{tabular}
\label{tab:sec2q3}
\end{table}
\begin{table}[!t]
\footnotesize
\centering
\caption{$A_q(n,i)$ in the quaternary case ($q=4$). The values are constructive (i.e., they correspond to actual codes found by an exhaustive search). Values in italics can also be taken from Proposition~\ref{prop:n1q}.}
\begin{tabular}{r|rrrrrrrrrrrr}
~~~$i$&1&2&3&4&5&6&7&8&9\\
$n$~~~& & & & & & & & & \\\hline
1   &\em 1&1&1& & & & & & \\
2   &\em 2&2&2&1&1&1& & & \\
3   &\em 3&3&3&2&1&1&1&1&1\\
4   &\em 4&4&4&3&-&-&-&-&-
\end{tabular}
\label{tab:sec2q4}
\end{table}
\fi

\subsection{Bounds on the Sizes of Generations}
The bounds from Section~\ref{sec:proofs} can also be extended to the $q$-ary case for laminar codes with $n=\left \lceil \frac{t}{q-1} \right \rceil$ and the size of each generation maximized assuming no knowledge of the previous generation.

\begin{proposition} \label{prop:1}
  For any $n\geq 2$, $q\geq 2$, and $2\leq i\leq n$, $A_q(n,i)\geq \min(A_q(n-1,i-1),A_q(n-1,i))$.
\end{proposition}

\begin{IEEEproof}
The proof follows the same lines as the proof of Proposition~\ref{prop:aqni_vs_aqn-1}, with the suitable partition $\mathcal{Y}'$ of $E_q(n,i)$ defined as the union for all 
$1\leq k\leq \min(A_q(n-1,i-1),A_q(n-1,i))$ 
of the codeword classes
\begin{equation} \notag
  \left(f_{\mathcal{Y}}(k).0\right)\,\cup\,\bigcup_{s=1}^{q-1}\left(f_{\mathcal{Z}}(k).s\right).
\end{equation}
\end{IEEEproof}

\begin{proposition} \label{prop:n1q}
  For any $n\geq 1$ and $q\geq 2$, $A_q(n,1)=n$.
\end{proposition}

\begin{IEEEproof}
Same proof as for Proposition~\ref{prop:aqn1}.
\end{IEEEproof}

\begin{proposition}
  \label{prop:aqn2}
  For $n\geq 1$ and $q\geq 3$, $A_q(n,2)\geq A_2(n,2)+1$.
\end{proposition}

\begin{IEEEproof}
  Consider a suitable partition  $\mathcal{Y}$ of $E_2(n,2)$ of cardinality $A_2(n,2)$. Now consider $\mathcal{Y}'=\mathcal{Y}\cup \{2\bolde_k^n\,|\,1\leq k\leq n\}$. The cardinality of $\mathcal{Y}'$ is $A_2(n,2)+1$, the words in its codeword classes have weight $2$, and they belong to $E_q(n,2)$. There is no collision since $\mathcal{Y}$ 
has no collision, and the words we add are not in $E_2(n,2)$.
\end{IEEEproof}

\begin{proposition}
  \label{prop:aq2^n2}
  For $n\geq 0$ and $q\geq 3$, $A_q(2^n,2)\geq 2^n$.
\end{proposition}

\begin{IEEEproof}
It follows from direct application of Propositions~\ref{prop:a22^n2} and \ref{prop:aqn2}.
\end{IEEEproof}

\begin{proposition} \label{prop:2}
  For $n\geq 0$ and $q\geq 3$, $A_q(2n+1,2)\geq 2n+1$.
\end{proposition}

\begin{IEEEproof}
  The idea is to consider a codeword class whose circular permutations do not overlap. For $n=3$, such a codeword class is $\{0002000,0010100,0100010,1000001\}$.
  
  Formally, let us consider the following codeword class $Y_0$ of $E_q(2n+1,2)$:
  \begin{equation} \notag
    Y_0 = \left\{\bolde_{n+1-k}+\bolde_{n+1+k} \,|\, 0\leq k\leq n\right\}.
  \end{equation}
  $Y_0$ covers $E_q(2n+1,1)$. If $\mathcal{Y}=\{Y_0,Y_1,\dots,Y_{2n}\}$ is the family of the circular permutations of $Y_0$, then $\mathcal{Y}$ is a suitable partition of $E_q(2n+1,2)$. Indeed, for a given right circular permutation of $(\bolde_{n+1-k}+\bolde_{n+1+k})$, $k$ can be identified as follows.
  \begin{itemize}
  \item The vector has a $2$ if and only if $k=0$.
  \item Otherwise, it has two $1$'s at indices $i_1$ and $i_2$ with $i_1<i_2$. If $i_2-i_1$ is even, $k=\frac{i_2-i_1}{2}$ and we have permuted $(\bolde_{n+1-k}+\bolde_{n+1+k})$ to the right $i_1-n-1+k$ times. If $i_2-i_1$ is odd, $k=\frac{2n+1+i_1-i_2}{2}$ and we have permuted $(\bolde_{n+1-k}+\bolde_{n+1+k})$ to the right $i_2-n-1+k$ times.
  \end{itemize}
  The cardinality of $\mathcal{Y}$ is $2n+1$, which is a lower bound on the maximum cardinality of a suitable partition of $E_q(2n+1,2)$.
\end{IEEEproof}

Finally, we remark that the lower bounds of Propositions~\ref{prop:1}, \ref{prop:aqn2}, \ref{prop:aq2^n2}, and \ref{prop:2} match the exact values of $A_q(n,i)$ from 
\ifonecolumn
\else
Tables \ref{tab:sec2q3} and \ref{tab:sec2q4} 
\fi
for several values of $(n,i)$.

\subsection{The Construction from Section~\ref{sec:construction}} \label{subsec:construction}
The construction of Section~\ref{sec:construction} can be extended to $q$-ary codes as follows.

\begin{theorem} \label{th:construction_qary}
Let $C$ be an $[n,t:M_1,\dots,M_t]_q$ synchronous $q$-ary code of WOM-rate $R$, and $C'$ an $[n',t':M_1',\dots,M_{t'}']_2$ synchronous binary code of WOM-rate $R'$, both not containing the all-zero codeword. Then there exists an $[nn',tt':M_1M_1',\dots,M_1M_{t'}',\dots,M_tM_1',\dots,M_tM_{t'}']_q$ synchronous $q$-ary code $C_1$ of WOM-rate $R_1=\frac{t'}{n'}R+\frac{t}{n}R'$.
\end{theorem}

\begin{IEEEproof}
  The proof that $C_1$ is a valid synchronous $q$-ary code is the same as in the binary case.
\end{IEEEproof}

Notice that the code $C'$ in the construction is still binary: the requirement is that $C$ and $C_1$ must have the same  alphabet size. Using a $q'$-ary code (with $q'>2$) instead of a binary code is also possible regardless of $C$ and $C_1$. When $C'$ is binary, the two values $0$ and $1$ will be matched, at each stage, to $2$ successive generations $p-1$ and $p$ of $C$. In the first stage they are matched to generations $0$ (i.e., empty memory) and $1$, then to generations $1$ and $2$, and so on. However, when $C'$ is $q'$-ary with $q'>2$, each stage has $q'$ possible values to match to $q'$ generations. For instance, if $q'=4$, the values $(0,1,2,3)$ will be matched to generations $(0,1,2,3)$ of $C$ at stage $1$, then to generations $(3,4,5,6)$ at stage $2$, generations $(6,7,8,9)$ at stage $3$, and so on. 

If a nonbinary code $C'$ is to be used, then either $C$ or $C'$ must have a suitable structure. The following conditions, for example, would ensure this.

\begin{itemize}
\item A first sufficient condition is that each write of $C'$ increases the sum of the values of its cells by exactly one. This prevents the following situation from happening. Consider the case where at the first generation of a nonbinary $C'$, a cell can go both from $0$ to $1$ and from $0$ to $2$ depending on which message we encode. Then, in the corresponding block, we will write a codeword of $C$ of either generation $1$ or generation $2$. When encoding a pair $(m,m')$ of messages, the number of messages among which we can choose $m$ therefore depends on $m'$, which means that the encoder cannot predict how much data it will be able to store at a given generation.
\item Another possible condition to avoid the above issue is that we choose a fixed-rate code $C$. In the previous example, if $M_1=M_2$, it does not matter if we do not know whether we will be using generation $1$ or generation $2$ of $C$; we have the same number of messages to choose from anyway.
\end{itemize}

As an example, a $[4,10:4,2,4,2,6,3,4,2,2,1]_4\sync$ code can be constructed in the following way. First, a $[2,4:2,2,3,3]_4\sync$ code can be made by merging together the last three generations of the $[2,6:2,2,2,1,1,1]_4\sync$ code displayed above in Section~\ref{sec:qary} by taking as the new set of codeword classes the union of the sets
of codeword classes of the three last generations, and reorganizing them, as explained for the binary case in Section~\ref{sec:w(imi)_disjoint}. Also, if the codeword classes are reorganized properly, then an additional codeword class $\{22\}$ can be added to the third generation, resulting in the following $[2,4:2,2,3,3]_4\sync$ code
\ifonecolumn
\else
   \begin{equation*} \notag
     \hspace{-.02\textwidth}
    \begin{array}{r|cccc}
                   & 1                       & 2   & 3         \\ \hline
      \calD_1^{-1} & \{01\}                 & \{10\}   &   -\\
      \calD_2^{-1} & \{11\}             & \{20,02\}    &  -\\
      \calD_3^{-1} & \{21,03\}             & \{12,30\} & \{22\}     \\
      \calD_4^{-1} & \{13,32\}             & \{31,23\} & \{33\}     \\
    \end{array}
  \end{equation*}%
\fi
of WOM-rate $2.5850$.\footnote{By adding a generation containing the all-zero codeword, we get a $[2,5:1,2,2,3,3]_4\sync$ code of the same WOM-rate, which is significantly higher than the corresponding \emph{worst-case} WOM-rate of the synchronous lattice-based code from \cite[Table I]{bha14}.} This is the example code of Example~\ref{ex:sync_not_laminar}. Obviously, a $[2,5:2,2,3,2,1]_4\sync$ code can be made by splitting the fourth generation into the two generations $\{ \{13,32\}, \{31,23\} \}$ and $\{33\}$. Finally, by using the construction of Theorem~\ref{th:construction_qary}, using the $[2,5:2,2,3,2,1]_4\sync$ code as $C$ and the $[2,2:2,1]_2\sync$ code from Example~\ref{sec:example} as $C'$, a $[4,10:4,2,4,2,6,3,4,2,2,1]_4\sync$ code of WOM-rate $3.5425$ can be constructed.

\begin{table*}[t]
\footnotesize
  \centering
  \caption{WOM-rates of binary decodable codes obtained by concatenating synchronous codes, with target code length $n=64$. The numbers in the parentheses (in columns three and five) are the rate losses (computed from (\ref{eq:Gbasic}) and (\ref{eq:Gsync}), respectively) in percent, while the rate loss reduction factor is their fraction.} 
\begin{tabular}{cccccc}
\toprule
 $\tnd$             & Rate of nondec.                       & Rate of dec.          & \multicolumn{2}{c}{With data}     & Rate loss        \\
                 & code from \cite[Table VI]{yaa12_1}                            & with no data          & Sync. code           & Rate        & reduction factor       \\
\otoprule
\multirow{2}*{4} & \multirow{2}*{1.8566}                  & \multirow{2}*{1.7696 (4.69\%)} & $[3,4:1,3,1,1]_2      $ & 1.7943 (3.35\%)  & 1.40           \\
                 &                                        &                       & $[5,4:1,5,3,6]_2      $ & 1.8130 (2.35\%)           &2.00  \\[0.5mm]
\multirow{1}*{5}               & \multirow{1}*{1.9689}                                 & \multirow{1}*{1.8458 (6.25\%)}                & $[4,5:1,4,3,1,1]_2$ & 1.9019  (3.41\%)            & 1.84 \\[0.5mm]
 6               & 2.1331                                 & 1.9665 (7.81\%)                & $[5,6:1,5,3,2,1,1]_2  $ & 2.0431 (4.22\%) & 1.85             \\[0.5mm]
\multirow{2}*{7} & \multirow{2}*{2.1723}                  & \multirow{2}*{1.9686 (9.38\%)} & $[6,7:1,6,5,3,1,1,1]_2$ & 2.0701 (4.71\%)      & 1.99       \\
                 &                                        &                       & $[8,7:1,8,4,6,3,4,2]_2$ & 2.0909 (3.75\%)        & 2.50     \\[0.5mm]
\midrule
\bottomrule

\end{tabular}
  \label{table:rates_64}
\end{table*}

\begin{table*}[t]
\footnotesize
  \centering
  \caption{WOM-rates of binary decodable codes obtained by concatenating synchronous codes, with target code length $n=256$. The numbers in the parentheses (in columns three and five) are the rate losses (computed from (\ref{eq:Gbasic}) and (\ref{eq:Gsync}), respectively) in percent, while the rate loss reduction factor is their fraction.}
\begin{tabular}{cccccc}
\toprule
 $\tnd$             & Rate of nondec.                       & Rate of dec.          & \multicolumn{2}{c}{With data}         & Rate loss   \\
                 & code from \cite[Table VI]{yaa12_1}                             & with no data          & Sync. code           & Rate            & reduction factor     \\
                 \otoprule
\multirow{2}*{4} & \multirow{2}*{1.8566}                  & \multirow{2}*{1.8348 (1.17\%)} & $[3,4:1,3,1,1]_2      $ & 1.8410 (0.84\%) & 1.40              \\
                 &                                        &                       & $[5,4:1,5,3,6]_2      $ & 1.8457 (0.59\%)         & 2.00     \\[0.5mm]
\multirow{1}*{5}               & \multirow{1}*{1.9689}                                 & \multirow{1}*{1.9381 (1.56\%)}                & $[4,5:1,4,3,1,1]_2$ & 1.9521 (0.85\%)     & 1.84        \\[0.5mm]
 6               & 2.1331                                 & 2.0914 (1.95\%)                & $[5,6:1,5,3,2,1,1]_2  $ & 2.1106 (1.05\%) & 1.85             \\[0.5mm]
\multirow{2}*{7} & \multirow{2}*{2.1723}                  & \multirow{2}*{2.1214 (2.34\%)} & $[6,7:1,6,5,3,1,1,1]_2$ & 2.1467 (1.18\%)    & 1.99        \\
                 &                                        &                       & $[8,7:1,8,4,6,3,4,2]_2$ & 2.1520 (0.94\%)     & 2.50        \\[0.5mm] 
\midrule
\bottomrule
\end{tabular}
  \label{table:rates_256}
\end{table*}

\section{Results and Comparison with the Standard Method}

\label{sec:results}


In this section, we use the synchronous codes derived in the previous sections to construct decodable codes from nondecodable ones as explained in Section~\ref{sec:MainIdea} (binary case) and Section~\ref{sec:qary} (nonbinary case). We compare the proposed method with the basic method that adds $\left\lceil \frac{\tnd-1}{q-1} \right\rceil$ cells containing no data. For this comparison, we consider two different target code lengths, $n=64$ and $n=256$. We then assume for each value of $n$ and for some specific values of $\tnd$, that there exists a $\tnd$-write code with WOM-rate equal to the best (i.e., of highest WOM-rate) codes from \cite{yaa12_1,nonbinarywomcodes}, and with length $\nnd=n-\nsync$, where $\nsync$ is the length of the synchronous code. Note that we do not use the actual code lengths at which these state-of-the-art WOM-rates are reached because they are very large \cite{discussion_kayser} and not explicitly stated in \cite{yaa12_1,nonbinarywomcodes}. However, this gives a meaningful comparison, since the rate loss with our approach (see (\ref{eq:Gsync})) is an increasing function of $\Rnondec$ when $n$, $\nsync$, and $R_{\rm sync}>0$ are fixed. Since no code (for any block length) of strictly higher WOM-rate than the ones reported in \cite{yaa12_1,nonbinarywomcodes} is (as far as we can tell) currently known, and considering a specific block length $\nnd$ will  likely reduce the WOM-rate of the best nondecodable code, the comparison is a sort of worst-case scenario for our approach.

The results for the binary case are reported in Tables \ref{table:rates_64} and \ref{table:rates_256}. We consider values for $\tnd$ between $4$ and $7$. The second column of each table reports the state-of-the-art WOM-rate of nondecodable codes, for each value of $\tnd$. The third column shows the WOM-rate that is obtained by appending $\tnd-1$ cells with no data to a length $\nnd=n-(\tnd-1)$ code with WOM-rate equal to the one reported in the second column. The next two columns show, for various synchronous codes, the WOM-rate that we obtain for the same target length. 
The $[3,4:1,3,1,1]_2\sync$, $[4,5:1,4,3,1,1]_2\sync$, $[5,6:1,5,3,2,1,1]_2\sync$, and $[6,7:1,6,5,3,1,1,1]_2\sync$ codes are obtained by adding to the codes $[3,3:3,1,1]_2\sync$, $[4,4:4,3,1,1]_2\sync$, $[5,5:5,3,2,1,1]_2\sync$, and $[6,6:6,5,3,1,1,1]_2\sync$ from Section~\ref{sec:w(imi)=i} a generation containing the all-zero codeword.\footnote{Note that from Table~\ref{table:bound_2}, $B(6,4)=2$, which implies that a $[6,6:6,5,3,2,1,1]_2\sync$ code may exist. However, we have not been able to identify such a code in a (nonexhaustive) computer search. The best code found was a $[6,6:6,5,3,1,1,1]_2\sync$ code.} The $[5,4:1,5,3,6]_2\sync$ code is obtained in a similar manner from the $[5,3:5,3,6]_2\sync$ code in Section~\ref{sec:w(imi)_disjoint}, and the $[8,7:1,8,4,6,3,4,2]_2\sync$ code is obtained by adding a generation with the all-zero codeword to the $[8,6:8,4,6,3,4,2]_2\sync$ code from the construction of Section~\ref{sec:construction}.

To better quantify the gains of the proposed approach, we have included in the tables the rate losses compared to the nondecodable code, and also their fraction (the rate loss reduction factor), which quantifies the reduction in rate loss of the proposed approach compared to the basic approach of appending $\tnd-1$ cells containing no data. For both lengths, our technique yields higher WOM-rates compared to just appending a block of $\tnd-1$ cells with no information. For instance, for $\tnd=7$ and $n=64$, the rate loss with the basic approach is as high as $9.38\%$. With the improved approach the rate loss is reduced to $3.75\%$, which is a reduction by a factor of $2.5$ (see the sixth column of Table~\ref{table:rates_64}). As can be seen from the tables, the rate loss of the basic approach grows with $\tnd$. In all cases we are able to demonstrate a rate loss reduction factor of $1.8$ to $2.5$ using our approach, which is significant.
Furthermore, the tabulated WOM-rates are (to the best of our knowledge) also higher than the best WOM-rates for binary multiple-write codes (and hence better than the WOM-rates of any directly decodable code) known prior to \cite{yaa12_1}, which justifies our approach.

\begin{table*}[t!]
\begin{minipage}{\linewidth}
\renewcommand\thefootnote{\thempfootnote}
\footnotesize
  \centering
  \caption{WOM-rates of quaternary ($q=4$) decodable codes obtained by concatenating  synchronous codes, with target code length $n=64$. The numbers in the parentheses (in columns three and five) are the rate losses (computed from (\ref{eq:Gbasic}) and (\ref{eq:Gsync}), respectively) in percent, while the rate loss reduction factor is their fraction.}
\begin{tabular}{cccccc}
\toprule
 $\tnd$             & Rate of nondec.                   & Rate of dec.          & \multicolumn{2}{c}{With data} & Rate loss               \\
                 & code                           & with no data          & Sync. code                       & Rate      & reduction factor  \\
\otoprule
5               & 3.9328  \cite{nonbinarywomcodes}                        & 3.8099  (3.13\%)              &  $[2,5:1,2,2,3,3]_4$                               &     3.8907   (1.07\%) & 2.92    \\[0.5mm]
6               & 4.2594  \cite{nonbinarywomcodes}                         & 4.1263  (3.13\%)              &  $[2,6:1,2,2,3,2,1]_4$                             &     4.1979   (1.44\%)  & 2.17    \\[0.5mm]
 \multirow{1}*{7}               & \multirow{1}*{4.3394 \cite{nonbinarywomcodes}}     & \multirow{1}*{4.2038 (3.13\%)}                & $[2,7:1,2,2,2,1,1,1]_4$            & 4.2507  (2.04\%) & 1.53   \\ [0.5mm]
8               & 4.5088\,\footnote{Obtained by applying Construction A from \cite{nonbinarywomcodes} to the WOM-rates from \cite[Table VI]{yaa12_1}.\\ \;\;\addtocounter{mpfootnote}{+1}\footnotemark\addtocounter{mpfootnote}{-1}Obtained by applying Construction A from \cite{nonbinarywomcodes} to the WOM-rates from the recursion for $\mathcal{R}'_t$ of Section~VI in \cite{yaa12_1}. }                             & 4.2975  (4.69\%)              &  $[3,8:1,3,3,3,2,1,1,3]_4$                               &     4.4121 (2.14\%) & 2.19      \\[0.5mm]
 9               & 4.5836\,\footnotemark[\value{mpfootnote}]               & 4.3687  (4.69\%)              & $[3,9:1,3,3,3,2,1,1,1,2]_4$            &   4.4743 (2.38\%) & 1.97  \\ [0.5mm]  
10               & 4.6932\,\footnotemark[\value{mpfootnote}]                & 4.4732 (4.69\%)               & $[3,10:1,3,3,3,2,1,1,1,1,1]_4$      & 4.5631   (2.77\%)    & 1.69 \\[0.5mm]
11               & 4.7193\,\addtocounter{mpfootnote}{+1}\footnotemark[\value{mpfootnote}]                & 4.4243 (6.25\%)               & $[4,11:1,4,2,4,2,6,3,4,2,2,1]_4$      & 4.6457  (1.56\%) & 4.01     \\[0.5mm]
\midrule
\bottomrule
\end{tabular}
  \label{table:q4rates_64}
\end{minipage}
\end{table*}

\begin{table*}[htbp]
\begin{minipage}{\linewidth}
\renewcommand\thefootnote{\thempfootnote}
\footnotesize
  \centering
  \caption{WOM-rates of quaternary ($q=4$) decodable codes obtained by concatenating  synchronous codes, with target code length $n=256$. The numbers in the parentheses (in columns three and five) are the rate losses (computed from (\ref{eq:Gbasic}) and (\ref{eq:Gsync}), respectively) in percent, while the rate loss reduction factor is their fraction.}
\begin{tabular}{cccccc}
\toprule
 $\tnd$             & Rate of nondec.                   & Rate of dec.          & \multicolumn{2}{c}{With data} & Rate loss               \\
                 & code                         & with no data          & Sync. code                       & Rate    & reduction factor   \\
\otoprule
5               & 3.9328  \cite{nonbinarywomcodes}                        & 3.9021   (0.78\%)             &  $[2,5:1,2,2,3,3]_4$                               &     3.9223 (0.27\%) & 2.92      \\[0.5mm]
6               & 4.2594  \cite{nonbinarywomcodes}                         & 4.2261 (0.78\%)               &  $[2,6:1,2,2,3,2,1]_4$                             &     4.2440  (0.36\%)   & 2.17     \\[0.5mm]
 \multirow{1}*{7}& \multirow{1}*{4.3394 \cite{nonbinarywomcodes}}     & \multirow{1}*{4.3055 (0.78\%)}                & $[2,7:1,2,2,2,1,1,1]_4$            & 4.3172   (0.51\%) & 1.53  \\ [0.5mm]
8               & 4.5088\,\footnote{Obtained by applying Construction A from \cite{nonbinarywomcodes} to the WOM-rates from \cite[Table VI]{yaa12_1}. \\ \;\;\addtocounter{mpfootnote}{+1}\footnotemark\addtocounter{mpfootnote}{-1}Obtained by applying Construction A from \cite{nonbinarywomcodes} to the WOM-rates from the recursion for $\mathcal{R}'_t$ of Section~VI in \cite{yaa12_1}.}                              & 4.4560 (1.17\%)                & $[3,8:1,3,3,3,2,1,1,3]_4$                               &    4.4846 (0.54\%) &2.19       \\[0.5mm]
 9               & 4.5836\,\footnotemark[\value{mpfootnote}]                & 4.5299  (1.17\%)              & $[3,9:1,3,3,3,2,1,1,1,2]_4$            &    4.5563 (0.60\%) & 1.97 \\[0.5mm]
10               & 4.6932\,\footnotemark[\value{mpfootnote}]                 & 4.6382 (1.17\%)               & $[3,10:1,3,3,3,2,1,1,1,1,1]_4$      & 4.6607 (0.69\%)   & 1.69   \\[0.5mm]
11               & 4.7193\,\addtocounter{mpfootnote}{+1}\footnotemark[\value{mpfootnote}]                & 4.6456 (1.56\%)               & $[4,11:1,4,2,4,2,6,3,4,2,2,1]_4$      &   4.7009  (0.39\%) & 4.01  \\[0.5mm]
\midrule
\bottomrule
\end{tabular}
  \label{table:q4rates_256}
\end{minipage}
\end{table*}
The results for the nonbinary case with $q=4$ are reported in  Tables \ref{table:q4rates_64} and \ref{table:q4rates_256} for $n=64$ and $n=256$, respectively. Here, we consider values for $\tnd$ between $5$ and $11$. As in the binary case, the second column of each table reports the state-of-the-art  WOM-rate of nondecodable quaternary codes, for each value of $\tnd$ that we consider. The third column shows the WOM-rate that would be obtained by appending $\left\lceil \frac{\tnd-1}{3} \right\rceil$ cells containing no data to a code of length $n- \left\lceil \frac{\tnd-1}{3} \right\rceil$ and WOM-rate equal to the one reported in the second column. Note that similar to the binary case, the codes that we have constructed in Section~\ref{sec:qary} can be extended by a single generation containing the all-zero codeword only. Thus, when we speak below about codes that are constructed in previous sections, we implicitly assume that they have been extended in this way. Now, the codes $[2,7:1,2,2,2,1,1,1]_4\sync$ and $[3,10:1,3,3,3,2,1,1,1,1,1]_4\sync$ are taken from Section~\ref{sec:qary} 
\ifonecolumn
\else
(the second and third rows of Table~\ref{tab:sec2q4}, respectively), 
\fi
the codes $[3,8:1,3,3,3,2,1,1,3]_4\sync$ and $[3,9:1,3,3,3,2,1,1,1,2]_4\sync$ are obtained by merging the last three (resp.\ two) generations of the $[3,10:1,3,3,3,2,1,1,1,1,1]_4\sync$ code, and the codes $[2,5:1,2,2,3,3]_4\sync$, $[2,6:1,2,2,3,2,1]_4\sync$, and $[4,11:1,4,2,4,2,6,3,4,2,2,1]_4\sync$ are taken from Section~\ref{subsec:construction}. 
 Note that as in the binary case our technique yields higher WOM-rates compared to just appending a block of $\left\lceil \frac{\tnd-1}{3} \right\rceil$ cells with no information, for both target lengths. Also, as in the binary case,   the rate loss of the basic approach grows with $\tnd$, and we demonstrate  a rate loss reduction by a factor between $1.5$ and $4.0$ in all cases considered, which is significant.

For the ternary case, to the best of our knowledge, no tables of the best possible WOM-rates have been presented in the literature. There are however constructions that can be used. See, for instance, \cite[Theorem 7]{yaa12_1} for constructing $q$-ary $2$-write codes. Here, we will use a construction from \cite{nonbinarywomcodes} (which was inspired by a similar idea proposed in \cite{hua09}) giving  a $q$-ary $2(q-1)$-write code of WOM-rate $(q-1) R_2$, where $R_2$ is the best possible WOM-rate of a $2$-write binary code. Thus, there exists a ternary $4$-write code of WOM-rate $(3-1) \cdot 1.4928 = 2.9856$ where the WOM-rate of the $2$-write code is taken from \cite[Table VI]{yaa12_1}. 
\ifonecolumn
\else
Now, from the second row of Table~\ref{tab:sec2q3}, we can see that 
\fi
there exists a $[2,3:2,2,2]_3\sync$ code (by merging the last two generations) that does not contain the all-zero codeword. Assuming a block length of $n=64$, our method gives a WOM-rate of $2.9392$, while the method of appending  $\left \lceil \frac{4-1}{3-1} \right \rceil = 2$ cells with no data gives a WOM-rate of only $2.8923$. This amounts to a rate loss reduction by a factor of $2.01$.



\section{Conclusion}
\label{sec:conclu}

In this paper, we proposed short synchronous WOM codes as a basic tool to make nondecodable codes decodable while preserving the WOM-rate as much as possible. We considered both binary and nonbinary codes, as well as the fixed-rate and the unrestricted-rate setups. We constructed short synchronous (laminar) codes for small values of $t$. We also proposed a construction method to build synchronous codes for higher values of $t$ by concatenating shorter synchronous codes. Compared to the construction by Rivest and Shamir, which considers $n=t$ with $t$ being a power of $2$, our construction is more general, since it lifts both constraints. Finally, we used the obtained synchronous codes to make some nondecodable codes decodable. Compared to the standard approach of appending cells containing no data, the proposed approach achieves a significant reduction of the rate loss for short-to-moderate block lengths.

\section*{Acknowledgment}

The authors wish to thank S.\ Kayser for valuable discussions and the anonymous reviewers for their valuable comments and suggestions that helped improve the presentation of the paper.


\balance



\begin{thebibliography}{10}
\providecommand{\url}[1]{#1}
\csname url@samestyle\endcsname
\providecommand{\newblock}{\relax}
\providecommand{\bibinfo}[2]{#2}
\providecommand{\BIBentrySTDinterwordspacing}{\spaceskip=0pt\relax}
\providecommand{\BIBentryALTinterwordstretchfactor}{4}
\providecommand{\BIBentryALTinterwordspacing}{\spaceskip=\fontdimen2\font plus
\BIBentryALTinterwordstretchfactor\fontdimen3\font minus
  \fontdimen4\font\relax}
\providecommand{\BIBforeignlanguage}[2]{{%
\expandafter\ifx\csname l@#1\endcsname\relax
\typeout{** WARNING: IEEEtran.bst: No hyphenation pattern has been}%
\typeout{** loaded for the language `#1'. Using the pattern for}%
\typeout{** the default language instead.}%
\else
\language=\csname l@#1\endcsname
\fi
#2}}
\providecommand{\BIBdecl}{\relax}
\BIBdecl

\bibitem{RivSha82}
R.~L. Rivest and A.~Shamir, ``How to reuse a ``write-once'' memory,''
  \emph{Information and Control}, vol.~55, no. 1-3, pp. 1--19, Oct./Nov./Dec.
  1982.

\bibitem{mer84}
F.~Merkx, ``Womcodes constructed with projective geometries,'' \emph{Traitement
  du Signal}, vol.~1, no. 2--2, pp. 227--231, 1984.

\bibitem{fia84}
A.~Fiat and A.~Shamir, ``Generalized ``write-once'' memories,'' \emph{IEEE
  Trans. Inf. Theory}, vol.~30, no.~3, pp. 470--480, May 1984.

\bibitem{capacitypermanentmemory}
C.~Heegard, ``On the capacity of permanent memory,'' \emph{IEEE Trans. Inf.
  Theory}, vol.~31, no.~1, pp. 34--42, Jan. 1985.

\bibitem{cho86}
G.~D. Cohen, P.~Godlewski, and F.~Merkx, ``Linear binary code for write-once
  memories,'' \emph{IEEE Trans. Inf. Theory}, vol.~32, no.~5, pp. 697--700,
  Sep. 1986.

\bibitem{zem91}
G.~Z\'{e}mor and G.~D. Cohen, ``Error-correcting {WOM}-codes,'' \emph{IEEE
  Trans. Inf. Theory}, vol.~37, no.~3, pp. 730--734, May 1991.

\bibitem{capacitygeneralizedWOM}
F.-W. Fu and A.~J.~H. Vinck, ``On the capacity of generalized write-once memory
  with state transitions described by an arbitrary directed acyclic graph,''
  \emph{IEEE Trans. Inf. Theory}, vol.~45, no.~1, pp. 308--313, Jan. 1999.

\bibitem{yaa10}
E.~Yaakobi, P.~H. Siegel, A.~Vardy, and J.~K. Wolf, ``Multiple error-correcting
  {WOM}-codes,'' in \emph{Proc. IEEE Int. Symp. Inf. Theory (ISIT)}, Austin,
  TX, Jun. 2010, pp. 1933--1937.

\bibitem{twowritewomcodes}
E.~Yaakobi, S.~Kayser, P.~H. Siegel, A.~Vardy, and J.~K. Wolf, ``Efficient
  two-write {WOM}-codes,'' in \emph{Proc. IEEE Inf. Theory Workshop (ITW)},
  Dublin, Ireland, Aug./Sep. 2010.

\bibitem{multiplewritewomcodes}
S.~Kayser, E.~Yaakobi, P.~H. Siegel, A.~Vardy, and J.~K. Wolf, ``Multiple-write
  {WOM}-codes,'' in \emph{Proc. 48th Annual Allerton Conf. Commun., Control,
  and Computing}, Monticello, IL, Sep./Oct. 2010, pp. 1062--1068.

\bibitem{nonbinarywomcodes}
R.~Gabrys, E.~Yaakobi, L.~Dolecek, P.~H. Siegel, A.~Vardy, and J.~K. Wolf,
  ``Non-binary {WOM}-codes for multilevel flash memories,'' in \emph{Proc. IEEE
  Inf. Theory Workshop (ITW)}, Paraty, Brazil, Oct. 2011, pp. 40--44.

\bibitem{yaa12_1}
E.~Yaakobi, S.~Kayser, P.~H. Siegel, A.~Vardy, and J.~K. Wolf, ``Codes for
  write-once memories,'' \emph{IEEE Trans. Inf. Theory}, vol.~58, no.~9, pp.
  5985--5999, Sep. 2012.

\bibitem{shp13}
A.~Shpilka, ``New constructions of {WOM} codes using the {W}ozencraft
  ensemble,'' \emph{IEEE Trans. Inf. Theory}, vol.~59, no.~7, pp. 4520--4529,
  Jul. 2013.

\bibitem{bha14}
A.~Bhatia, M.~Qin, A.~R. Iyengar, B.~M. Kurkoski, and P.~H. Siegel,
  ``Lattice-based {WOM} codes for multilevel flash memories,'' \emph{IEEE J.
  Sel. Areas Commun.}, vol.~32, no.~5, pp. 933--945, May 2014.

\bibitem{bha12}
A.~Bhatia, A.~R. Iyengar, and P.~H. Siegel, ``Multilevel $2$-cell $t$-write
  codes,'' in \emph{Proc. IEEE Inf. Theory Workshop (ITW)}, Lausanne,
  Switzerland, Sep. 2012, pp. 247--251.

\bibitem{jia07}
A.~Jiang, ``On the generalization of error-correcting {WOM} codes,'' in
  \emph{Proc. IEEE Int. Symp. Inf. Theory (ISIT)}, Nice, France, Jun. 2007, pp.
  1391--1395.

\bibitem{jia08}
A.~Jiang and J.~Bruck, ``Joint coding for flash memory storage,'' in
  \emph{Proc. IEEE Int. Symp. Inf. Theory (ISIT)}, Toronto, ON, Canada, Jul.
  2008, pp. 1741--1745.

\bibitem{mah09}
H.~Mahdavifar, P.~H. Siegel, A.~Vardy, J.~K. Wolf, and E.~Yaakobi, ``A nearly
  optimal construction of flash codes,'' in \emph{Proc. IEEE Int. Symp. Inf.
  Theory (ISIT)}, Seoul, Korea, Jun./Jul. 2009, pp. 1239--1243.

\bibitem{discussion_kayser}
S. Kayser, private communication.

\bibitem{hua09}
Q.~Huang, S.~Lin, and K.~A.~S. Abdel-Ghaffar, ``Error-correcting codes for
  flash coding,'' \emph{IEEE Trans. Inf. Theory}, vol.~57, no.~9, pp.
  6097--6108, Sep. 2011.

\end{thebibliography}

\end{document}